\documentclass[useAMS,usenatbib]{mn2e}

\newcommand{\etal}{et~al.\ }

\def\msun{\,{\rm M_\odot}}

\def\spose#1{\hbox to 0pt{#1\hss}}
\def\lta{\mathrel{\spose{\lower 3pt\hbox{$\mathchar"218$}} \raise 2.0pt\hbox{$\mathchar"13C$}}}
\def\gta{\mathrel{\spose{\lower 3pt\hbox{$\mathchar"218$}} \raise 2.0pt\hbox{$\mathchar"13E$}}}

\newcommand\beq{\begin{equation}}
\newcommand\eeq{\end{equation}}

\title[Early high-$\sigma$ peaks in present-day CDM haloes]
{The distribution and kinematics of early high-$\sigma$ peaks in present-day
haloes: implications for rare objects and old stellar populations}
\author[J\"urg Diemand, Piero Madau $\&$ Ben Moore]
{J\"urg Diemand$^{1,2}$\thanks{email: diemand@ucolick.org.}, Piero Madau$^1$, 
Ben Moore$^2$\\
1. Department of Astronomy and Astrophysics, University of
California, Santa Cruz, CA 95064.\\
2. Institute for Theoretical Physics, University of Z\"urich,
Winterthurerstrasse 190 ,CH-8057 Z\"urich, Switzerland.}

\begin{document}
\pagerange{\pageref{firstpage}--\pageref{lastpage}} \pubyear{2005}
\maketitle
\label{firstpage}  
\begin{abstract}
We show that the hierarchical assembly of cold dark matter (CDM) haloes preserves the 
memory of the initial conditions. Using N-body cosmological simulations, we demonstrate
that the present-day spatial distribution and kinematics of objects that formed 
within {\it early} ($z\gta 10$) protogalactic systems (old stars, satellite galaxies, 
globular clusters, 
massive black holes, etc.) depends mostly on the rarity of the peak of the primordial 
density field which they originally belonged to.
Only for objects forming at lower redshifts the exact formation site within the 
progenitor halo (e.g. whether near the center or in an extended disk) becomes important.
In present-day haloes, material from the rarer early peaks is more 
centrally concentrated and falls off more steeply with radius
compared to the overall mass distribution, it has a lower velocity dispersion, moves
on more radial orbits, and has a more elongated shape. Population II stars that formed
within protogalactic haloes collapsing from $\ge 2.5\sigma$ fluctuations would follow
today a $r^{-3.5}$ density profile with a half-light radius of 17 kpc and 
a velocity anisotropy that increases from isotropic in the inner regions
to nearly radial at the halo edge. This agrees well with the radial velocity dispersion 
profile of Galaxy halo stars from Battaglia et al. (2005) and 
with the anisotropic orbits of nearby halo stars.
\end{abstract}
\begin{keywords}
methods: N-body simulations -- 
Galaxy: halo -- kinematics and dynamics -- 
galaxies: formation -- haloes -- star clusters
\end{keywords}
 
\section{Introduction}

In a Universe where cold dark matter (CDM) dominates structure formation, the haloes 
of galaxies and clusters are 
assembled via the hierarchical merging and accretion of smaller progenitors (e.g.
\citealt{Lacey1993}). This process causes structures to relax violently to a 
new equilibrium by
redistributing energy among the collisionless mass components.
Early stars formed in these progenitors behave as a collisionless
system just like the dark matter particles in their host haloes, and they undergo the same
dynamical processes during subsequent mergers and the buildup of larger systems 
like massive galaxies or clusters. It is of crucial importance in galaxy formation
studies to explore the efficiency of 
the mixing process and see if any spatial or kinematical
signatures exist in material that collapses at different epochs and within 
peaks of the primordial Gaussian density field of different rarity.

In this paper, we use a suite of high-resolution cosmological N-body simulations to  
analize the distribution and kinematics within present-day galaxy haloes
of dark matter particles that originally belonged to selected branches of the 
merger tree. These properties are particularly relevant for the baryonic tracers of 
early CDM structures, e.g. the old stellar halo which may have originated from the 
disruption at high redshift of numerous dwarf protogalaxies \citep{Bullock2000},
the old halo globular clusters, and also giant ellipticals \citep*{Gao2004}. 
The end product of the 
entire merger tree is a triaxial cuspy dark matter halo (\citealt{Dubinski1991};
\citealt{NFW}; \citealt*{Moore1999pro}; \citealt{Diemand2005}): a small fraction of 
early progenitor systems survive the merging process and end up as dark matter 
substructures \citep{Ghigna1998}. 
Since rare, early haloes are strongly biased towards overdense regions (e.g. 
\citealt{Cole1989}; \citealt{Sheth1999}), i.e. towards the centers of larger scale 
fluctuations that have not collapsed yet, we might expect that 
material originating from the earliest branches 
of the merger tree is today much more centrally concentrated than the overall halo.
Indeed, a ``non-linear'' peak biasing has been discussed previously by several authors  
(\citealt{Moore1998}; \citealt{White2000}; \citealt{Moore2001}). 

Here we show that the distribution and kinematics of ``old material'' 
within present-day galaxy haloes depends 
primarily on the rareness of the peaks of the primordial density fluctuation
field it originally belonged to. Specifically, today's properties of 
objects that formed in old rare density peaks above $\nu\sigma(M,z)$ [where 
$\sigma(M,z)$ is the {\it rms} fluctuation in the
density field linearly extrapolated to redshift $z$
smoothed with a top-hat filter of mass $M$. A ``one $\sigma$ peak''
corresponds to the characteristic mass 
``$M_*(z)$'' and $\sigma(M_*,z) \simeq 1.69$],
depend largely on $\nu$ and not on the particular  
values of $z$ and $M$.  Such centrally concentrated components 
are isotropic in the inner part, just like the host galaxy haloes, but 
rapidly become more radially anisotropic further out.
The plan of the paper is as follows. In \S~2 we describe the numerical simulations 
and how to define and trace high-$\sigma$  particle subsets. In \S~3
we analyse the present-day distribution of these subsets.  We derive a simple
empirical fitting formula for $\rho(r,\nu)$, the mass density profile of 
all progenitors above $\nu\sigma$, which approximates the results of our N-body
cosmological simulations for $1<\nu<4$. \S~4 discusses the implications
of our findings for old stellar populations. We argue that such centrally concentrated 
components are predicted to be isotropic in the inner part, just like the host galaxy 
halo, but to rapidly become more radially anisotropic further out. This is 
quantitative with the radial velocity dispersion 
profile of Galaxy halo stars from \citet{Battaglia2005} and 
with the anisotropic orbits of nearby halo stars 
($\beta \simeq 0.5$, \citealt{Chiba2000}).
Finally, we present our conclusions in \S~5.

\section{Method}\label{Method}

We identify collapsed high-$\sigma$ peaks at different epochs within high-resolution 
cosmological N-body simulations, mark them, and analyse the distribution and 
kinematics of this material at redshift zero. Details about
the simulations and halo finding method are given in the next two subsections.

\subsection{Simulations} \label{sim}

The simulations have been performed using PKDGRAV, a parallel N-body treecode
written by Stadel and Quinn \citep{Stadel2001}: cosmological and numerical 
parameters are the same as in \citet{Diemand2004pro}. The present-day haloes that 
we analyse are labeled ``D12'' (cluster) and ``G0'' to ``G3'' (galaxies). We also 
analyse an additional smaller galaxy halo ($M_{\rm vir} =10^{11} \msun$), labeled
``G4''. 

\begin{table}\label{haloes}

\caption{Present-day properties of the six simulated dark matter haloes.
The columns give halo name, spline softening length, number
of particles within the virial radius, virial mass, virial radius, peak
circular velocity, and radius to the peak of the circular velocity curve.
}
\begin{tabular}{l | c | c | c | c | c | c }
  \hline
  Halo&$\epsilon_0$&$N_{\rm vir}$&
$M_{\rm vir}$&$r_{\rm vir}$&$V_{c,{\rm max}}$&$r_{V_{c,{\rm max}}}$ \\
 &[kpc]&[$10^6$]&$[10^{12} \msun]$&[kpc]&[km/s]&[kpc] \\
 \hline
 $D12$&1.8&14.0 & $305$ & 1743 & 958 & 645\\
 $G0$&0.27&1.7 & $1.01$ & 260 & 160 & 52.2\\
 $G1$&0.27&1.9 & $1.12$ & 268 & 162 & 51.3\\
 $G2$&0.27&3.8 & $2.21$ & 337 & 190 & 94.5\\
 $G3$&0.27&2.6 & $1.54$ & 299 & 180 & 45.1\\
 $G4$&0.27&0.25 & $0.144$ & 138 & 96.4 & 15.0\\
  \hline   
\end{tabular}
\end{table}

\begin{figure*}
\vskip 9 truein
\includegraphics{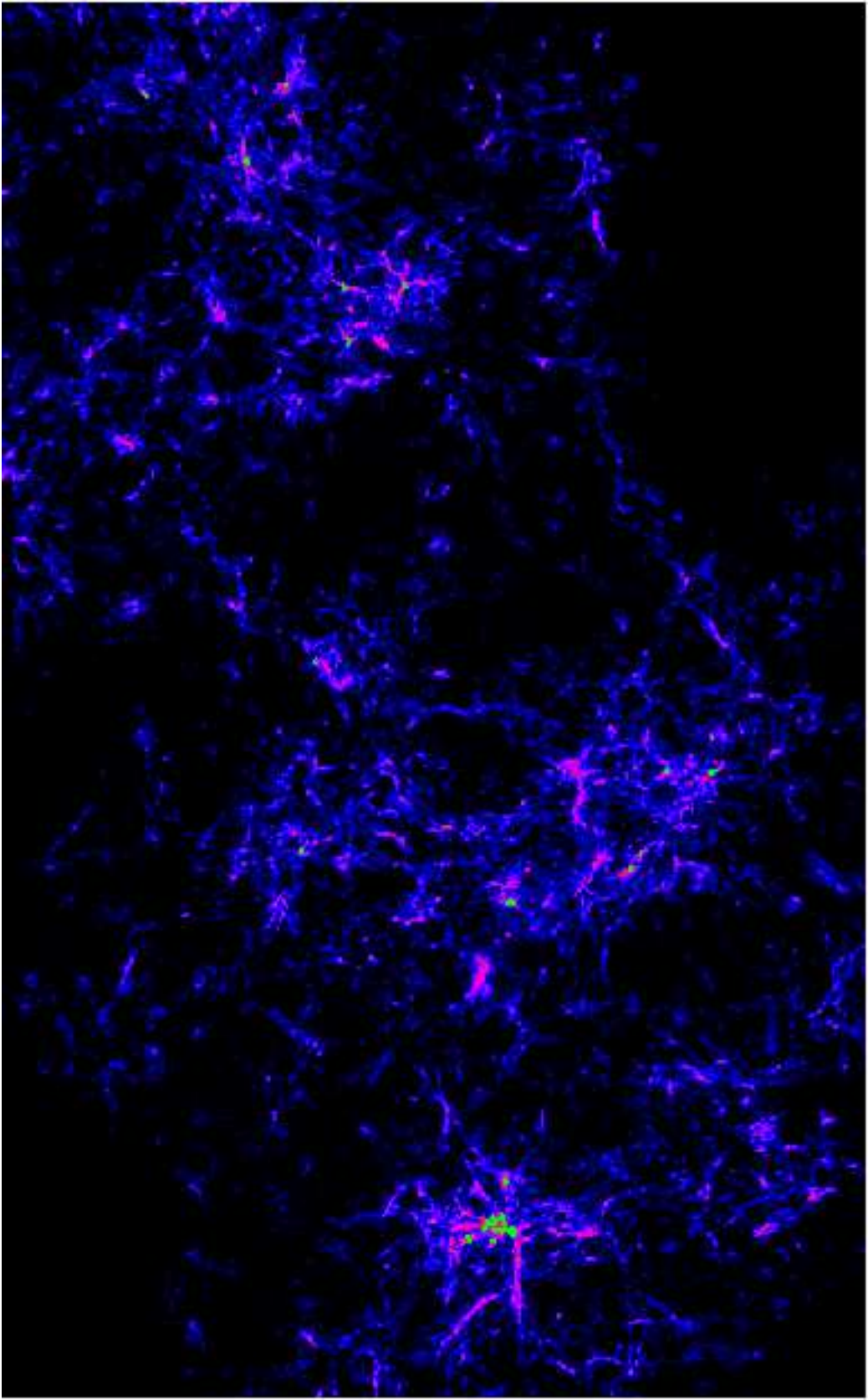}
\caption{Density map of the high-resolution region of run G at $z=13.7$.
FOF groups more massive than $4.9 \times 10^7 \msun$ 
(=84 $m_{\rm DM}$) are marked in green color.}
\label{4}
\end{figure*}

Figures 1 and 2 show the density field of the region containing the G0-G4 galaxy haloes.
Most of the material belonging to high-$\sigma$ peaks in the high-resolution 
region can be found in the four most massive galaxy haloes, while smaller structures tend 
not to have any progenitors which meet the selection criteria. G4 is a special, 
early-forming small galaxy halo (with a high concentration of $c=18$) that contains 
similar fractions of high-$\sigma$ material as the
more massive galaxies. From Figure \ref{4} it is clear that G4 is not
a representative system for its mass range.
All ``G'' haloes form in the one region of $\sim$ (10 Mpc)$^3$ that was selected from 
a (90 Mpc)$^3$ box and re-simulated at high resolution. The region has a mean density of
0.5 times the background density at $z=0$.
The properties of the six haloes are given in Table 1. 

\begin{figure*}
\vskip 9 truein
\includegraphics{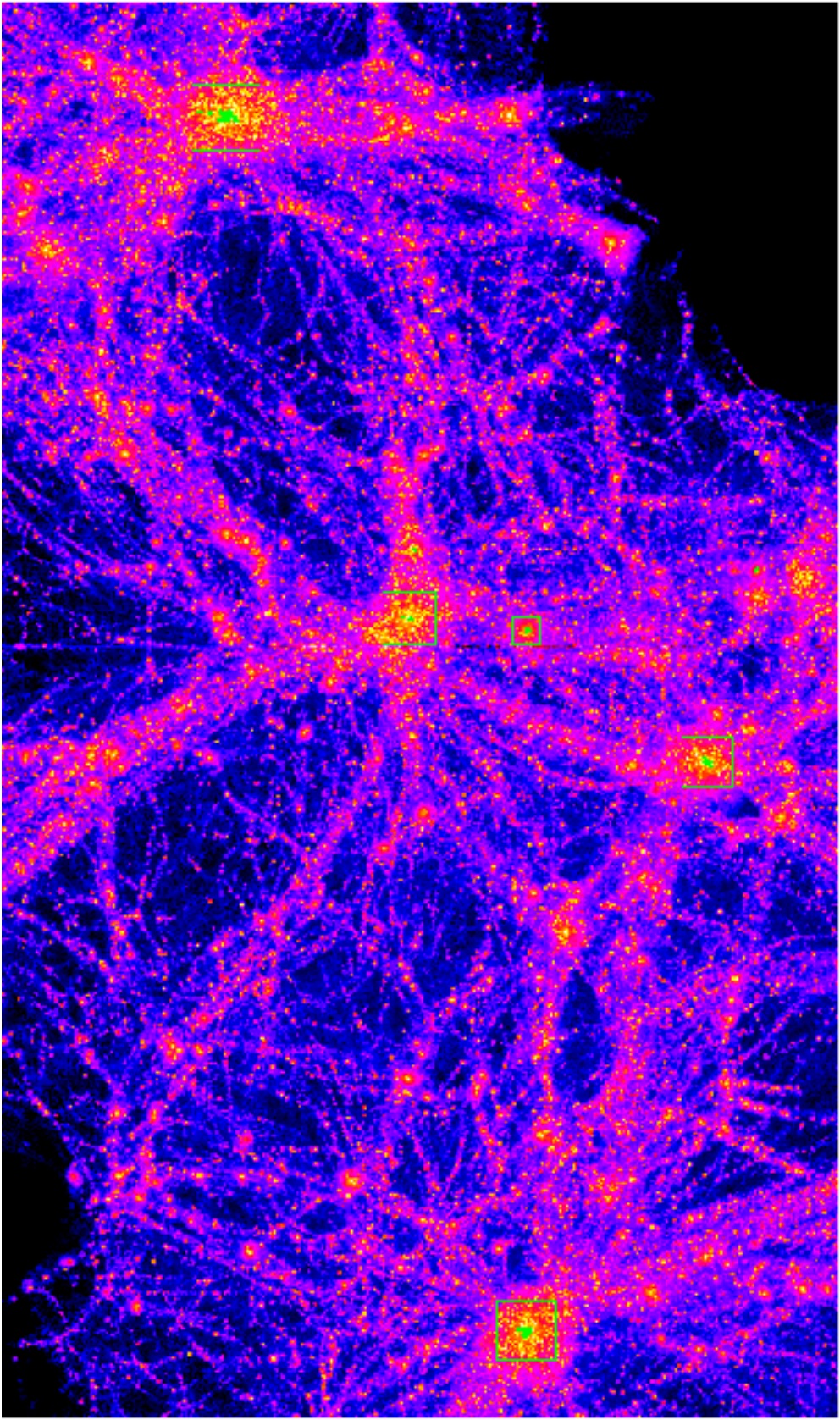}
\caption{Density map of the high-resolution region of run G at $z=0$.
The marked particles (in {\it green}) were selected at $z=13.7$ in groups 
above $4.9 \times 10^7 \msun$, see Fig. \ref{4}. The 
squares enclose the virial radii of the five galaxy haloes analysed (G0 to G4).} 
\label{200}
\end{figure*}

\begin{figure*}
\vskip 5.5 truein
\includegraphics{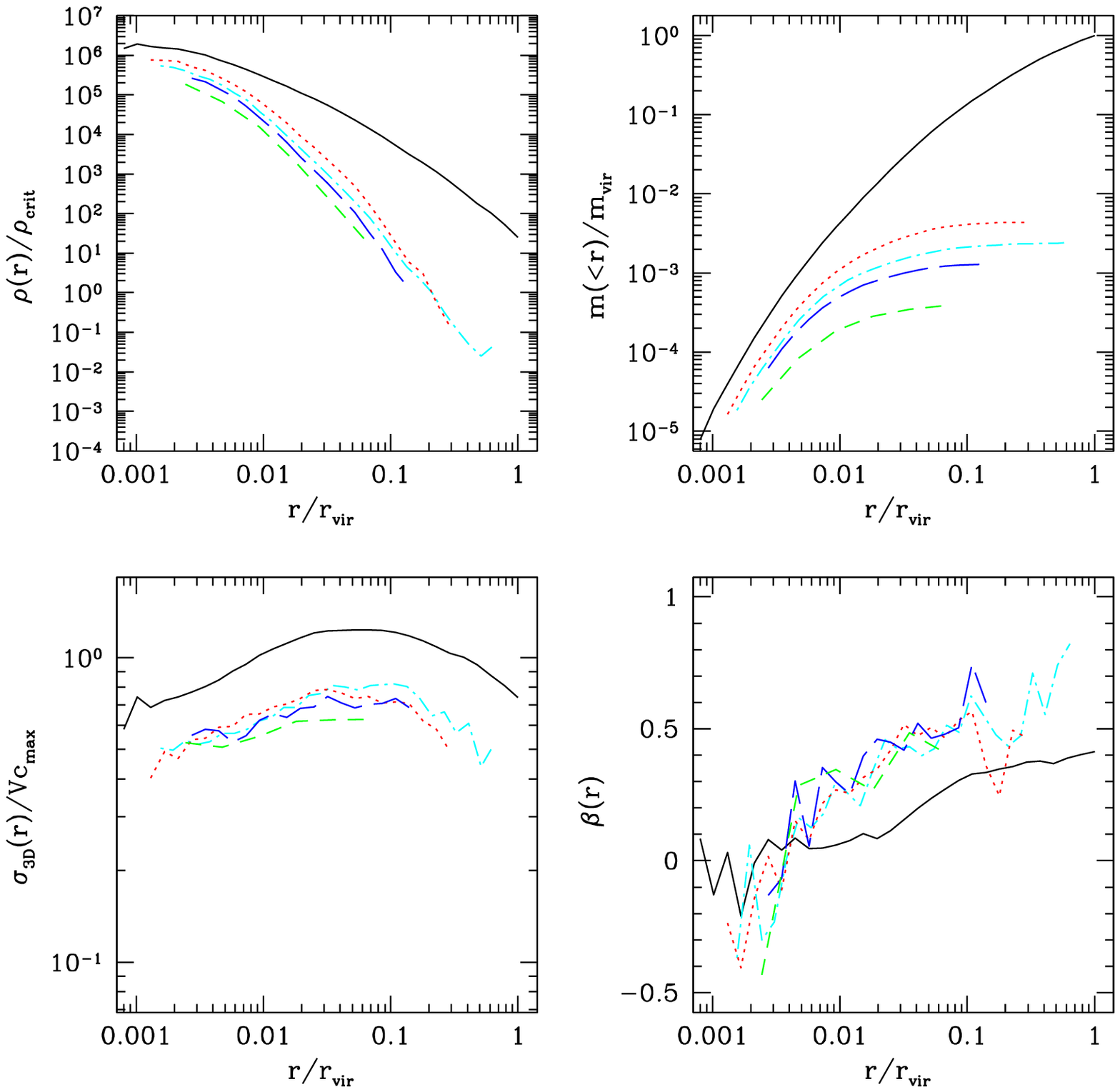}
\caption{The present-day distribution and kinematics 
of dark matter particles selected at $z=10.5$, 
averaged over four Milky Way-size simulated haloes.
We marked the 52 most massive groups found with FOF
using different linking lengths: 
$b=0.164$ ({\it dotted lines}), $b=0.164/2$ ({\it long-dashed lines}) and
$b=0.164/3$ ({\it short-dashed lines}). Particles selected
at the same redshift with overdensities above 1000 have similar present-day
distributions ({\it dash-dotted lines}).
For comparison we also plot the same quantities 
using all particles from the final structure ({\it solid curves}).
} 
\label{sel}
\end{figure*}

The cosmological parameters are $(\Omega_m,\Omega_{\Lambda},\sigma_8,h)
=(0.268, 0.732, 0.7, 0.71)$. The value of $\sigma_8=0.9$
given in \citet{Diemand2004pro} is not correct: we found that, due to a 
mistake in the normalization, our initial conditions have less power than intended.
According to linear theory, mass fluctuations grow proportional to the 
scale factor in a flat $\Omega_m=1$ Universe, which
is a good approximation to the adopted $\Lambda$CDM Universe
at $z>2$. Therefore the scale factor at collapse of a
given halo in our simulations would be 0.78 times smaller in a $\sigma_8=0.9$ model.
Throughout this paper, together with the collapse redshifts given for our 
low-$\sigma_8$ simulations, we will also state the rescaled collapse redshifts in a 
$\sigma_8=0.9$ Universe using the exact growth function for
a $(\Omega_m, \Omega_\Lambda)=(0.3, 0.7)$ cosmology
(see table \ref{sigma}). Using a  $\sigma_8=0.9$ resimulation
of the ``D'' cluster we confirm that the $z=0$ results (in units of the scale 
radius) are consistent with those of the $\sigma_8=0.7$ simulation when we use 
the $z_{0.9}$ outputs of the $\sigma_8=0.9$ run to select progenitor haloes. 

\begin{table*}
\centering
\begin{minipage}{140mm}
\caption{Properties of 4, 3.5, 3, 2 and 1$\sigma$ fluctuations collapsing at different
output redshifts. Here $N_{\rm groups}$ is the number of groups found with mass $>M$, and 
$f$ is the fraction of the overall mass in the high-resolution region which is inside 
these groups (the total mass in the high-resolution region is $3.6\times 10^{13}\msun$
in run G, and  $1.6\times 10^{15}\msun$ in run D12). 
The fractions differ from the average value, $f={\rm erfc}(\nu/\sqrt{2})$, 
as the high-resolution regions analysed are relatively small 
and are not representative patches of the Universe. We did not use outputs when 
the mass scale lies below ten times the particle mass, or when the number of 
groups is smaller than two per present-day halo.
In the cluster run (D12) we have only analysed one representative case
for 1$\sigma$ and 2$\sigma$ progenitors.}
\label{sigma}
\begin{tabular}{l | c | c | c || c | c || c | c }
  \hline
  $\nu$&$z$&$z_{0.9}$&$M_{\rm min}$&
$N_{\rm groups}\in G$&$f \in G$&
$N_{\rm groups}\in D12$&$f \in D12$\\
 & & &$[\msun]$& & & & \\
 \hline
4 & 10.5 & 13.8 & $ 3.4\times 10^{9}$&-&-& 24&$1.2 \times 10^{-4}$\\
4 & 8.7 & 11.5 &  $ 1.8\times 10^{10}$&-&-& 14&$2.9\times 10^{-4}$\\
4 & 7.4 & 9.8 & $ 6.2\times 10^{10}$&-&-& 8&$5.8\times 10^{-4}$\\
4 & 5.7 & 7.7 & $ 3.6\times 10^{11}$&-&-&3&$1.2\times 10^{-3}$\\
 \hline
3.5 & 16.3 & 21.2 & $ 5.8\times 10^{6}$&72&$2.5\times 10^{-5}$&-&-\\
3.5 & 13.7 & 17.9 & $ 4.9\times 10^{7}$&35&$1.3\times 10^{-4}$&-&-\\
3.5 & 10.5 & 13.8 & $ 8.4\times 10^{8}$&(5)& & 236&$3.4\times 10^{-4}$\\
3.5 & 8.7 & 11.5 &  $ 5.2\times 10^{9}$&-&-& 126&$1.0\times 10^{-3}$\\
3.5 & 7.4 & 9.8 & $ 1.9\times 10^{10}$&-&-& 60&$1.7\times 10^{-3}$\\
3.5 & 4.4 & 6.0 & $ 6.6\times 10^{11}$&-&-& 8&$8.5\times 10^{-3}$\\
 \hline
3 & 13.7 & 17.9 & $ 5.9\times 10^{6}$&670&$3.0\times 10^{-4}$&-&-\\
3 & 10.5 & 13.8 & $ 2.2\times 10^{8}$&52&$8.7\times 10^{-4}$&1659&$7.1\times 10^{-4}$\\
3 & 8.7 & 11.5 & $ 1.0\times 10^{9}$&23&$1.6\times 10^{-3}$&1205&$2.5\times 10^{-3}$\\
3 & 7.4 & 9.8 & $ 4.3\times 10^{9}$&8&$2.0\times 10^{-3}$&501&$4.3\times 10^{-3}$\\
3 & 4.4 & 6.0 & $ 2.3\times 10^{11}$&-&-&27&$1.4\times 10^{-2}$\\
 \hline
2.5 & 10.5 & 13.8 & $ 1.8\times 10^{7}$&1426&$2.5\times 10^{-3}$& &\\
2.5 & 8.7 & 11.5 & $ 1.0\times 10^{8}$&558&$5.1\times 10^{-3}$& &\\
\hline
2 & 8.7 & 11.5 & $ 7.6\times 10^{6}$&12740&$1.1\times 10^{-2}$&-&-\\
2 & 7.4 & 9.8 & $ 4.8\times 10^{7}$&2657&$1.5\times 10^{-2}$&-&-\\
2 & 4.4 & 6.0 & $ 6.3\times 10^{9}$&43&$2.1\times 10^{-2}$&2459&$2.1\times 10^{-2}$\\
 \hline
1 & 2.4 & 3.5 & $4.8\times 10^{8}$&1413&0.14&-&-\\
1 & 1.6 & 2.4 & $7.4\times 10^{9}$&108&0.13&4258&$0.24$\\
1 & 1.1 & 1.8 & $3.9\times 10^{10}$&24&0.11&-&-\\
1 & 0.8 & 1.5 & $1.0\times 10^{11}$&9&0.11&-&-\\
 \hline  
\end{tabular}
\end{minipage}
\end{table*}

\subsection{Tracing progenitor material}
\label{region}

Since baryonic objects (i.e. early protogalaxies, first stars) will form 
in the inner parts of their host gravitational potential, it is important to assess how different 
their $z=0$ distribution is when only the inner part of a progenitor
halo is traced. We will therefore compare the final distribution of
material marked within the central regions of early haloes with the entire
marked progenitor. We will show in this section that
there is no difference (except of course in the traced mass fraction) if the
traced objects are {\it early} ($z\gta 10$) protogalaxies. 

Progenitor haloes are identified using the friends-of-friend algorithm (FOF)
\citep{Davis1985}. For simplicity we use a fixed linking length
$b=0.164$ at all output times. Our results are not sensitive to 
this choice. For example at redshift 10.5 we
found 52 haloes more massive than $4.35 \times 10^9 \msun$ in the high 
resolution region of the galaxy simulation. 
When we use a linking length two or three times smaller and mark
the linked material in the same 52 groups we obviously find a smaller
fraction of marked matter but very similar density profile shapes and
kinematics (see Fig. \ref{sel}). We also selected the cores
of haloes at the same redshift by marking all particles which
have a local density (calculated from an SPH kernel over 32 nearest neighbours)
$10^3$ times higher than the mean matter density and obtained similar
$z=0$ distributions. This demonstrates that at redshift zero, particles
originating from the cores of high-$\sigma$ haloes are
distributed in the same way as all the progenitor halo material, and justifies 
our choice of using all halo particles as tracers for baryonic objects which 
would likely form in the very center of their hosts. A large number of tracers give two
important advantages. First, it allows us to reliably estimate density,
velocity dispersion, and anisotropy profiles. Second, it makes
the results more robust against numerical effects: since two-body relaxation
completely changes the orbits of many individual particles, even in 
high-resolution cosmological simulations \citep{Diemand2004rel}, using
only one most bound particle as a tracer for (say) a Population III remnant is
not a safe choice.

\begin{figure*}
\vskip 5.5 truein
\includegraphics{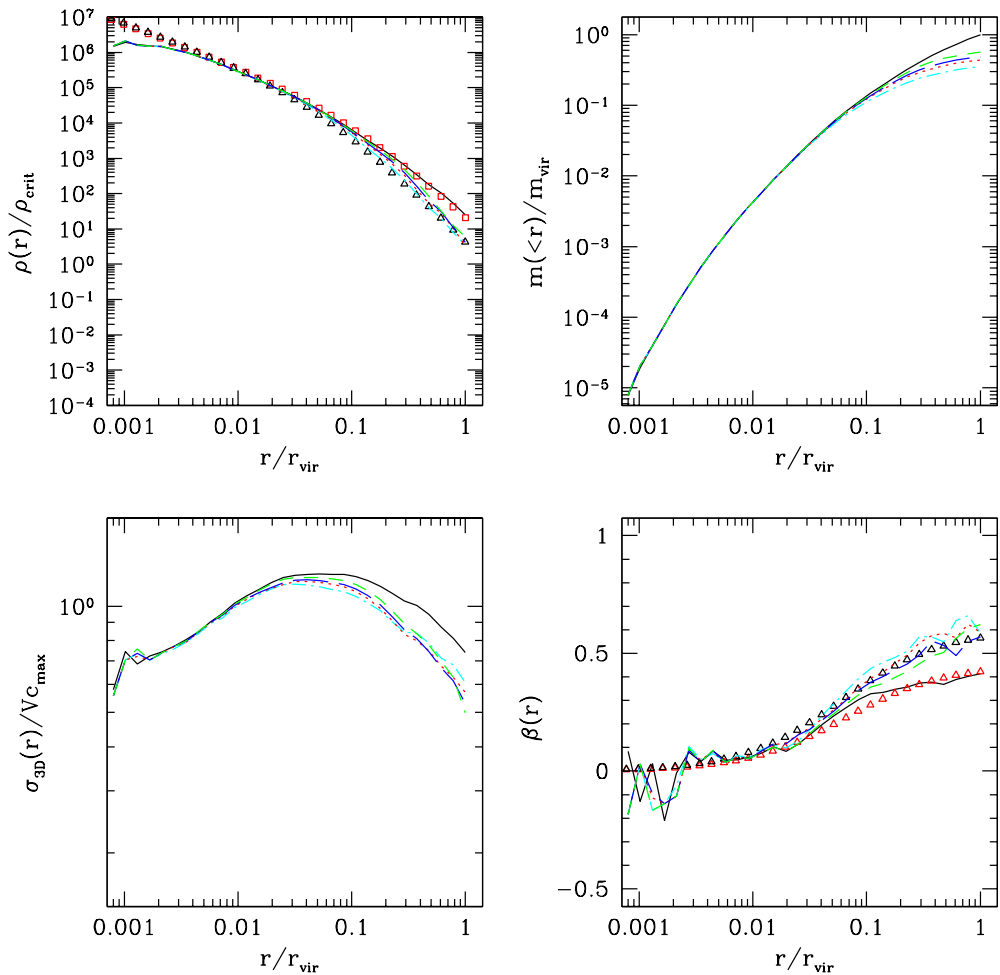}
\caption{The present-day distribution and kinematics of peaks above 1$\sigma$
averaged over four parent Milky Way-size galaxies. Here and in subsequent plots, the 
same-$\sigma$ peaks selected at the highest and up to the next three output redshifts
(given in Table 2) are plotted as dash-dotted, dotted, long-dashed and short-dashed lines,
respectively. For comparison we also plot the same quantities using all 
particles ({\it solid curves}). The triangles in the upper left and lower right 
panels show the empirical fitting functions in (\ref{emp}) and (\ref{bemp}).
} 
\label{avg1sig}
\end{figure*}

\begin{figure*}
\vskip 5.5 truein
\includegraphics{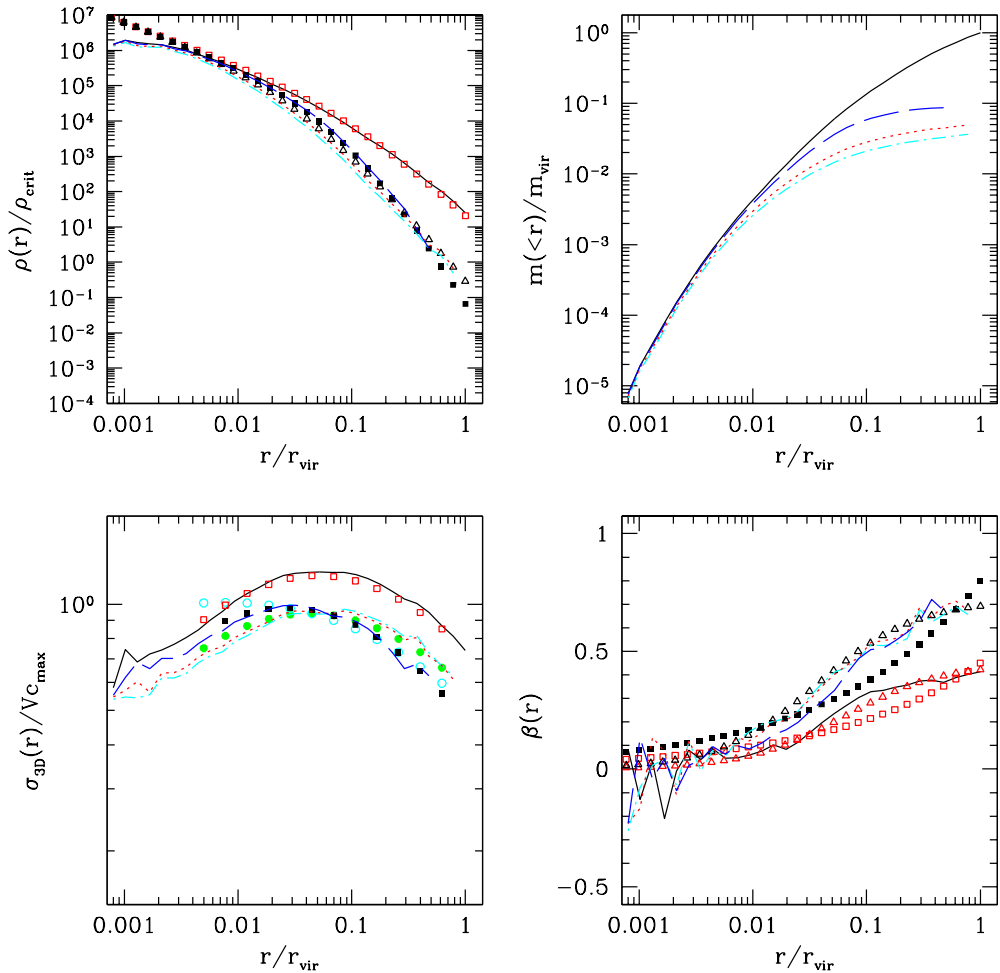}
\caption{Distribution and kinematics of peaks above 2$\sigma$ in present-day 
galaxy-size haloes. Line styles are as in Fig. \ref{avg1sig}.
The squares in the lower left panel are calculated from the Jeans
equation (\ref{je}) using approximations for $\rho(r)$ and 
$\beta(r)$ as input (plotted with squares in the upper left and
lower right panels). Dispersions from an isotropic model ({\it filled circles})
and a model with constant radial anisotropy $\beta(r)=0.5$ ({\it open circles}) 
do not fit the corresponding data ({\it long-dashed lines}) very well.} 
\label{avg2sig}
\end{figure*}

We have shown above that the final distribution and kinematics of particles from high-redshift, 
low-mass progenitor haloes is insensitive to their original location within the host potential.
The orbital energy associated with the merging of these small subunits into larger systems determines
the present-day distribution and dominates the relatively small differences
in the orbital and potential energies of particles within their hosts. 
This is not true at lower redshifts and for progenitor masses closer to the mass of 
the $z=0$ parent halo. For example, progenitors above $1\sigma$ marked at $z=1.6$ 
(i.e. with masses above $7.4\times10^9 \msun$) show only mild radial and velocity bias when
the entire FOF groups are marked (Fig. \ref{avg1sig}): the difference from the total dark 
matter distribution becomes much larger, however, if we select within these haloes only 
particles with overdensity greater than $10^4$, as this material is more 
concentrated, 
colder, and on more radial orbits (just like in the low-mass 2$\sigma$ selection, 
see Fig. \ref{avg2sig}).

\section{Distribution and kinematics of high-$\sigma$ peaks} \label{distribution}

We use the linear growth approximation to structure formation to calculate the masses of 4, 3.5, 3, 2 and 
1$\sigma$ fluctuations collapsing at a given redshift. The redshifts of the simulation outputs 
and the halo masses corresponding to these fluctuations are given in Table \ref{sigma}. 
These values are for the cosmological model we have simulated $(\Omega_m,\Omega_{\Lambda},\sigma_8,h)
=(0.268, 0.732, 0.7, 0.71)$. In a $\sigma_8=0.9$ the same fluctuations would collapse earlier 
(at $z_{0.9}$, see \S~\ref{sim}).

\begin{figure*}
\vskip 5.5 truein
\includegraphics{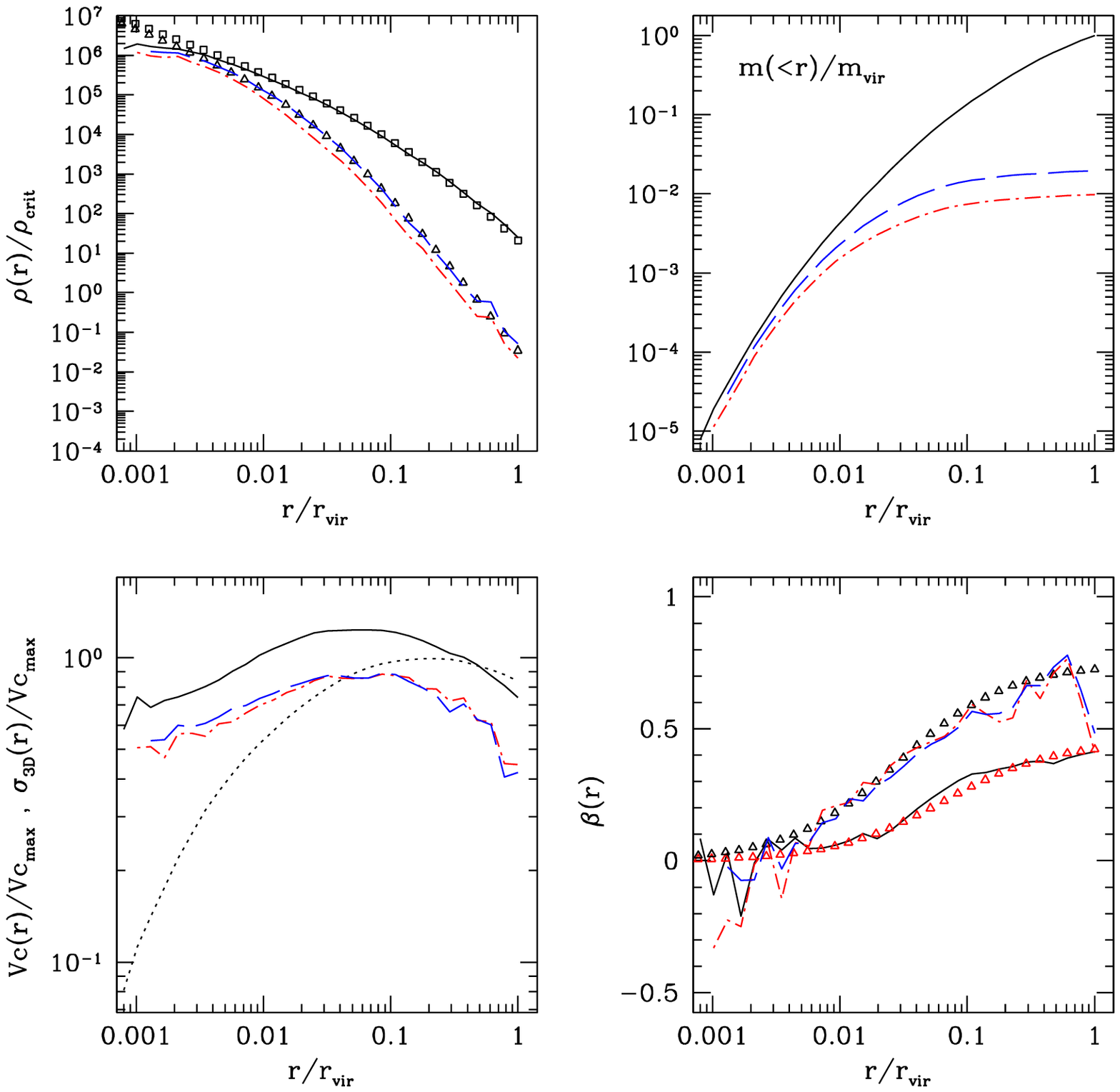}
\caption{Distribution of peaks above 2.5$\sigma$ in present-day galaxy-size haloes.
Line styles are as in Fig. \ref{avg1sig}.
The distribution resembles that of the Milky Way stellar halo.  
In the solar neighborhood ($r\simeq 0.03 r_{\rm vir}$),  the 3D
velocity dispersion is of order the circular velocity [$V_c(r)/V_{c,{\rm max}}$ 
plotted as dotted line in the lower left panel] and there is a clear radial anisotropy, 
with $\beta\simeq 0.4$ (see also Fig. \ref{MWhalo}).
} 
\label{avg2.5sig}
\end{figure*}

\subsection{Radial distribution} \label{rdistribution}

High-$\sigma$ material is strongly biased toward the center
of present-day haloes (see upper panels of Figs. \ref{avg1sig} to \ref{dallsig}),
with the rarer peaks showing stronger bias. Figures \ref{avg1sig}
to \ref{avg3.5sig} show different (M,z) selections corresponding to the same
value of $\nu$. While the shapes of these profiles are 
similar for a given $\nu$, the normalisation (or traced mass fraction)
generally grows with increasing mass threshold and to lower redshifts.
The density profiles are averaged over the four galaxy haloes of similar mass.
The scatter from halo to halo in their total density profiles 
and that of their subsets is relatively small.
But there are substantial halo-to-halo variations in the anisotropy
parameter $\beta(r)$, which we discuss in \S~\ref{kin}.

Mass densities are fitted with a general $\alpha\beta\gamma$-profile
that asymptotes to a central cusp $\rho(r) \propto r^{-\gamma}$:  
\begin{equation}\label{abc}
\rho(r,\nu) = \frac{\rho_s}{(r/r_{\nu})^{\gamma}[1 + (r/r_{\nu})^{\alpha}]^{(\beta-
\gamma)/\alpha} }.
\end{equation}
For comparison the NFW \citep*{NFW} profile has $(\alpha, \beta, \gamma) = (1,3,1)$, 
while the \citet*{Moore1999pro} profile has $(\alpha, \beta, \gamma) = (1.5,3,1.5)$. 
We fix $\alpha=1$ and the inner slope to $\gamma=1.2$, which is the best-fit slope for the 
$D12$ cluster when resolved at very high resolution ($m_{\rm DM}=3.0\times 10^5 \msun$)
\citep*{Diemand2005}.
We fit the entire dark halo using an outer slope of $\beta=3$ to determine
the scale radius $r_s = r_{vir}/c$, where $c$ is the concentration. To approximate
the high-$\sigma$ subset profiles we use a smaller scale radius $r_{\nu} \equiv r_s/f_{\nu}$
(corresponding to a higher concentration $c_{\nu} \equiv f_{\nu}c$)
and also a steeper outer slope $\beta_{\nu}$.
The $f_{\nu}$ and $\beta_{\nu}$ values used in the plots are calculated with simple
empirical formulae which approximately parameterise the entire range of
profiles, i.e. peaks above 1 to 4$\sigma$ and haloes ranging from a low concentration 
($c=4.5$) cluster halo to a small, $c=17$ galaxy halo:
\begin{equation}\label{emp}
r_{\nu} \equiv r_s/f_{\nu} \; , \;\;\; 
f_{\nu} = \exp(\nu/2) \; , \;\;\;
\beta_{\nu} = 3 + 0.26\nu^{1.6} \;\;.
\end{equation}
The values for the 1 to 4$\sigma$ peaks are given in Table \ref{sigmapro}, and 
the profiles are plotted in the upper left panels of Figures \ref{avg1sig} to \ref{dallsig} 
with open triangles. The fits are just approximate but they are reproduce well at least one 
profile for each of the $\nu$ values. Other parameters and functional forms could fit some of the 
data better.

For easier comparison with extragalactic observational data of old
stellar or globular cluster populations we also fit our high-$\sigma$ subset
profiles with a deprojected $R^{1/4}$ law, 
using the accurate numerical approximation of \citet{Marquez2001}:
\begin{eqnarray}\label{RQ}
\rho(r) &=& \rho_0 (r/R_e)^{-p} e^{-b(r/R_e)^{1/4}} \;,\\
p &=& 1.0 - 0.6097(1/4) + 0.05563(1/4)^2\;,\nonumber \\
b &=& 1.9992\times4-0.3271\;. \nonumber
\end{eqnarray}
The best-fit effective radii $R_e$ are given in Table \ref{sigmapro}. Like the 
$r_{\nu}$ values they scale with $r_s$ and not with the virial radius.

\begin{table}
\centering
\caption{Density profile parameters of 4, 3.5, 3, 2.5, 2, and 1$\sigma$ fluctuations in
parent haloes at $z=0$. The profiles are approximated
with the general function in eq. (\ref{abc}) using 
$(\alpha, \beta, \gamma)=(1,\beta_{\nu},1.2)$ and 
a concentration $c_{\nu}$ which is higher than the halo concentration
$c$ by some factor $f_{\nu}$, i.e. $c_{\nu} \equiv f_{\nu} c$.
The outer profile slopes $\beta_{\nu}$ and concentration factors $f_{\nu}$ 
given here are calculated from the empirical formulae in (\ref{emp}). These 
profiles are plotted in the corresponding figures as triangles. $R_e$ is 
the effective radius of the deprojected R$^{1/4}$ model (eq. \ref{RQ}) and
$r_{1/2}$ is the average half mass radius of the high-$\sigma$ material
at $z=0$. Both $R_e$ and $r_{1/2}$ 
are given in units of the scale radius of the entire dark matter
halo ($r_{\rm vir}/c$).
Values in brackets are for the cluster halo D12 which has a concentration of $c=4.5$.
For comparison, the bottom line gives the same parameters for the entire halo.
} 
\label{sigmapro}
\begin{tabular}{l | c | c | c | c }
\hline
$\nu$&$f_{\nu}$&$\beta_{\nu}$&$R_e$&$r_{1/2}$\\
\hline
4 &7.39&5.39&(0.067)&(0.1)\\
3.5&5.76&4.93&0.077&0.12(0.18)\\
3 &4.48&4.51&0.14&0.28(0.28)\\
2.5&3.49&4.13&0.32&0.51 \\
2 &2.72&3.79&0.71&0.70(0.83)\\
1 &1.65&3.26&2.9&1.8(1.2)\\
\hline  
halo& 1.0 & 3.0 &8.3& 3.5(1.8)\\
\hline  
\end{tabular}
\end{table}

\subsection{Kinematics}
\label{kin}

The lower left panels of Figures \ref{avg1sig} to \ref{dallsig} show the three
dimensional velocity dispersion profiles $\sigma_{\rm 3D}^2=\sigma_r^2 + \sigma_t^2$,
where $\sigma_r$ and $\sigma_t$ are the radial and the tangential velocity dispersions
($\sigma_t^2=\sigma_{\theta}^2 + \sigma_{\phi}^2$).
The lower right panels show the anisotropy parameter $\beta = 1 - 0.5 \sigma_t^2 / \sigma_r^2$
as a function of radius. The high-$\sigma$ subsets are clearly slower and on more radial
orbits than the entire dark matter component at the same radius. As with the density profiles,
the differences depend on $\nu$, i.e. the highest
$\sigma$ peaks have the lowest $\sigma_{\rm 3D}$ and the largest $\beta$ values.

Like the density profiles also the velocity dispersions are averaged over 
the four galaxy-size parent haloes. There is little halo-to-halo scatter 
in $\rho(r)$ and $\sigma_{\rm 3D}(r)$, but there is substantial scatter 
in the anisotropy parameter $\beta(r)$, especially near the virial radius. 
We have checked that the radial bias exists in each data set and is not just on
average. The variation in $\beta$ over the four galaxies is about 0.1 within 10\% 
of the virial radius, for both the parent halo and the progenitor subsets. Further out the 
scatter becomes much larger: particles in the parent halo (subsets) have a total spread of 0.45 
(0.35) near $r_{\rm vir}$.

\subsubsection{Jeans equations}

The origin of the smaller velocity dispersions of the subsets
can be traced to the fact that high-$\sigma$ peaks form
closer to the main progenitor of the present-day parent halo, hence
they join the system with small infall velocities and at early times when 
its potential well is much shallower. Since they have been part of the parent 
halo for a long time, they are likely in dynamical
equilibrium with the host, i.e. their density and velocity profiles should 
be a stationary solution to the Jeans equation in the potential, $\Psi(r)$, 
of the parent halo. Setting time derivatives to zero and 
assuming spherical symmetry, the Jeans equation can be written as \citep{BT}
\begin{equation}\label{je}
\frac{d}{dr} \left( \rho\sigma_r^2 \right) + \frac{2\beta}{r}
\rho\sigma_r^2  + \rho \frac{d\Psi}{dr}= 0 \;\;.
\end{equation}
Approximating the anisotropy parameter as in \cite{Diemand2004pro}, 
\begin{equation}\label{b}
\beta(r) = \beta_{\rm vir} (r/r_{\rm vir})^{1/3},
\end{equation}
equation (\ref{je}) admits the solution
\begin{equation}\label{sol}
\rho(r)\sigma_r^2(r) = G e^{-6\beta(r)} 
\int_r^{\infty}  e^{6\beta(y)} y^{-2}\rho(y) M(y) \; dy \; .
\end{equation}
Figure \ref{avg2sig} depicts this solution for 2$\sigma$ material.
We use equation (\ref{abc}) with $(\alpha, \beta_{\nu}, \gamma) = (1,5.4,1.2)$ and 
$r_{\nu}=r_s=r_{\rm vir}/10$ to approximate the
density profile (see long-dashed lines in Fig. \ref{avg2sig}), and 
equation (\ref{b}) with $\beta_{\rm vir}=0.8$ to approximate the anisotropy
profile $\beta(r)$. The radial velocity dispersion calculated from (\ref{sol})
is converted to $\sigma_{\rm 3D}$ and plotted (filled squares) in Figure \ref{avg2sig}.
It is very close to the measured velocity dispersions which confirms 
our expectation that high-$\sigma$ material is a more concentrated and 
colder subset of particles in dynamical equilibrium within the total dark
matter potential. This is just the opposite situation to the one of 
surviving CDM subhaloes, which are a more extended, hot subset 
in equilibrium with the host potential \citep*{Diemand2004sub}. 
We also calculated an isotropic [$\beta(r)=0$] and a constant $\beta(r)=0.5$ model
(see van den Bosch et al. 2004 for the corresponding solution of eq. \ref{je}) but
both cases do not to reproduce the measured $\sigma_{\rm 3D}(r)$ very well 
(see circles vs. long-dashed line in the lower left panel of Fig. \ref{avg2sig}).

As a consistency check we also solve eq. (\ref{sol}) for the entire dark 
matter halo, using $(\alpha, \beta_{\nu}, \gamma) = (1,3,1.2)$ and 
$\beta_{\rm vir}=0.45$ to approximate
the density and anisotropy profiles. The resulting $\sigma_{\rm 3D}$ is 
plotted (open squares) in the lower left panel of Figure \ref{avg2sig}.
The agreement with the measured dispersion is very good.
As expected our average system from four isolated and relaxed galaxy haloes is 
very close to a stationary equilibrium solution.

\subsubsection{A fitting function for $\beta(r,\nu)$}

Equation (\ref{b}) is not a very good approximation to the average
anisotropy $\beta(r)$ of the four G0-G3 parent galaxies. The function 
used by \citet{Mamon2005}
\begin{equation}\label{bmamon}
\beta(r) = \beta_a \frac{r}{r+ r_a}
\end{equation}
fits the data quite well for $\beta_a =0.45$ and $r_a=0.065$. For the 
high-$\sigma$ particles the radial anisotropy is larger and sets in further inside.
This behavior is approximated by simply
using the scale radii of the high-$\sigma$ density profiles (\ref{emp})
instead of $r_a=0.065$, and an amplitude $\beta_a$ that grows with $\nu$:
\begin{eqnarray}\label{bemp}
\beta(r, \nu) &=& \beta_{a\nu} \frac{r}{r+ r_{a\nu}} \;, \;\;\;
r_{a\nu} \equiv r_s/f_{\nu} \; , \nonumber \\
f_{\nu} &=& \exp(\nu/2) \; , \;\;\;
\beta_{a\nu}= 1 - 0.4 \nu^{-0.5} \;\;.
\end{eqnarray}
These anisotropy profiles are plotted with open triangles in the lower right
panels of Figures \ref{avg1sig} to \ref{avg3.5sig}. They also approximate 
the inner $\beta(r)$ of the haloes D12 and G4. Outside
10\% of the virial radius both the cluster D12 and the small galaxy G4
deviate substantially from the average $\beta(r)$ found for G0-G3.

\subsubsection{Density slope-velocity anisotropy relation}\label{slope-beta}

\citet{Hansen2004} have recently found a relation between the logarithmic slope 
of a density profile and the its velocity anisotropy:   
$\beta = \eta_1 - \eta_2 \; (d \log \rho / d \log r)$, with
$\eta_1$ ranging from -0.45 to 0.05 and $\eta_2$ from 0.1 to 0.35.
This relation approximates values measured in a variety of equilibrium   
N-body and SPH systems, including CDM haloes. It is interesting to ask
whether a similar relation also exist for our high-$\sigma$ subsets of present-day
galaxy haloes. By combining the fitting functions for $\rho(r,\nu)$
(eq. \ref{emp}) and $\beta(r,\nu)$ (eq. \ref{bemp}) one indeed finds 
the same simple relation,  
\begin{equation}\label{bvsslope}
\beta(d \log \rho / d \log r) = - \eta_2 (\gamma + d \log \rho / d \log r ) \;\;,
\end{equation}
where $\gamma=1.2$ is the slope of the inner density cusp ($\rho \propto r^{-1.2}$).
The coefficient $\eta_2$ depends only mildly on $\nu$.
Parent CDM haloes and low-$\sigma$ subsets have $\eta_2 \simeq 0.3$, while 
high-$\sigma$ subsets have smaller values, $\eta_2$=(0.29, 0.28, 0.23, 0.19) 
for $\nu$=(1, 2, 3, 4). The scatter around these relations is about 0.1 
when $\beta < 0.5$ (i.e. in the inner regions), and is larger (up to 0.4)
in the outer parts. A larger, more representative sample 
of CDM haloes should be studied to quantify more accurately the exact slope and
scatter of this important relation. The fact that even the high-$\sigma$ 
subsets follow a nearly universal relation between velocity anisotropy and density slope 
supports the existence of a fundamental connection between the two quantities
\citep{Hansen2004}. This relation breaks the mass-anisotropy degeneracy present 
when one uses the line-of-sight velocities of extragalactic stellar 
halo objects (such as globular clusters) to infer total dynamical masses.

\subsection{Shape of high-$\sigma$ material}

\begin{table*}
\centering
\begin{minipage}{180mm}
\caption{Mean fractional mass in present-day parent haloes of different
size ($2\times 10^{11}\,\msun$, $10^{12}\,\msun$, $10^{13}\,\msun$, and $10^{14}\,
\msun$) contributed by a fixed progenitor minimum mass/redshift 
($M_{\rm min}, z$) selection. Only parents hosting at least one 
selected progenitor are included when 
computing average values and scatter. The number of parent haloes (out of ten)
with selected progenitors is given in square brackets.} 
\label{sigmafr}
\begin{tabular}{l||ccc||ccc}
\hline
$M_{\rm min}, z$ &$10^{10}\msun$, 7.0&$10^{10}\msun$, 4.3&$10^{10}\msun$, 3.1
&$10^{11}\msun$, 4.3&$10^{11}\msun$, 3.1&$10^{11}\msun$, 0.8\\
 $\nu$&3.2&2.1&1.6&2.75&2.07&1.0\\
\hline
$2\times 10^{11}\msun$&0.0 [0]&0.049 [3]&0.158$\pm$0.075 [5]&0.0 [0]&0.0
[0]&0.567$\pm$0.071[6]\\
$10^{12}\msun$&0.0 [0]&0.047$\pm$0.035 [8]&0.149$\pm$0.068 [10]&0.0
[0]&0.123 [3]&0.598$\pm$0.074[10]\\
$10^{13}\msun$& 0.0033$\pm$0.0028 [8]&0.077$\pm$0.028
[10]&0.183$\pm$0.040[10]&0.036$\pm$0.025 [9]&
0.111$\pm$0.045 [10]&0.539$\pm$0.077 [10]\\
$10^{14}\msun$&0.0037$\pm$ 0.0021 [10]&0.072$\pm$0.014 [10]&0.179$\pm$0.022 [10]
&0.032$\pm$0.011 [10]&0.116$\pm$0.023 [10]&0.549$\pm$0.053 [10]\\
\hline
\end{tabular}
\end{minipage}
\end{table*}

\begin{figure}
\vskip 3.5 truein
\includegraphics{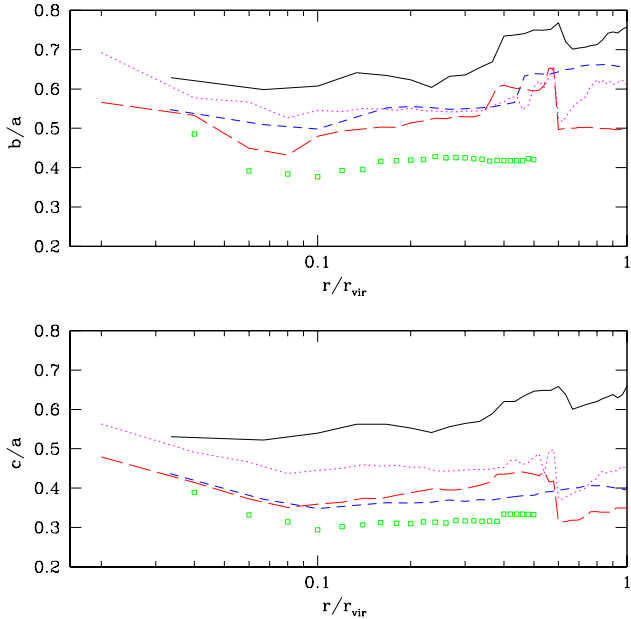}
\label{avgfofmom}
\caption{Shape profiles of high-$\sigma$ subsets at $z=0$ averaged over 
four galaxy haloes. The shapes are plotted for 3.5$\sigma$  
(groups with virial mass $\ge 4.9 \times10^{7} M_\odot$ at $z=13.7$, {\it squares}),
3$\sigma$ [$\ge 5.9 \times10^{6} M_\odot$ at $z=13.7$ ({\it long-dashed lines)} 
and $\ge 2.2 \times10^{8} M_\odot$ at $z=10.5$ ({\it short-dashed lines})]
and 2$\sigma$ material 
($\ge 4.8 \times10^{7} M_\odot$ at $z=7.5$, {\it dotted lines}).
The axis ratios of the entire dark matter halo are plotted with solid lines.}
\end{figure}

The axis ratios $a/c$ and $b/c$ are calculated with TIPSY
\footnote{Available at: http://www-hpcc.astro.washington.edu/.}
using the technique described in \citet*{Katz1991}. First the
inertia tensors of the particles within a sphere of a given radius
is calculated and diagonalized, then the same is done iteratively for particles 
within the triaxial shape found in the previous step, until the
procedure converges. Note that the shape at a given radius depends
on all particles within this radius, but for density profiles less steep than
$\rho(r) \propto r^{-4}$ the outer particles dominate over the
inner ones in the contribution to the inertia tensor.

The average axis ratios of the four galaxy haloes G0-G3 at the virial 
radius ($a/c \simeq 0.6,  a/b \simeq 0.7$)
are close to the mean values of large samples of haloes 
(\citealt*{Jing2002}; \citealt{Faltenbacher2002}; \citealt*{Bailin2005}).
Figure \ref{avgfofmom} shows the average shapes of 2, 3, and 3.5$\sigma$
material at the present epoch. Both axis ratios become smaller for higher-$\sigma$ material
at all radii.(This is also true for each of the four galaxies individually, 
not just for the average.) This result is consistent with the finding of 
\citet{Jing2002} that the average axis ratios decrease with redshift,
although our 2$\sigma$ haloes formed at higher redshift and lower virial mass 
than the range probed by \citet{Jing2002}. The origin of 
the extremely prolate shape of the high-$\sigma$ subsets 
($a:b:c\simeq 3:1:1$ for the 3.5$\sigma$ material) could be a series of
correlated head-on (i.e. low-angular momentum) mergers along
a filament aligned with the long axis \citep{Moore2004}.

For the total dark matter halo the axis orientations change with
radius, as found in \citet{Jing2002}, and a similar behavior is observed for 
the high-$\sigma$ subsets. The highest
$\sigma$ subsets show a slightly better alignment
which is simply explained with the fact that inner particles 
dominate the shape calculation due to the very high concentrations of these subsets.
Between the different subsets the axes are generally well aligned only
when they have many particles in common, i.e. in the inner parts.
As the fraction of high-$\sigma$ material drops with increasing radius,
the alignment with the shape of the whole halo becomes worse.

\subsection{Mass fraction of high-$\sigma$ material}\label{fraction}

It is well known that high-$\sigma$ peaks at early times are biased towards 
overdense regions where larger haloes will form later 
(see Figs. \ref{4} and \ref{200}). But how large is the mass fraction of such peaks
within present-day haloes, and how does this fraction depend on
the mass of the $z=0$ parent host? Do haloes which correspond to higher-$\sigma$ peaks
today (like massive clusters) contain a larger fraction of early high-$\sigma$ material?
This question is not well defined, since the fraction
of particles from peaks above 2$\sigma$ in a present-day galaxy halo 
grows when the peaks are selected with a larger mass 
threshold at lower redshift (see upper right panel of Fig. \ref{avg2sig}).
Here we use fixed mass threshold/redshift pairs
to select progenitors and compare the mass fractions they contribute
to parent haloes of different sizes. The selection of a fixed progenitor mass 
may be motivated in studies of old stellar populations if, for example:
a) Population III stars form in ``minihaloes'' above a {\it molecular cooling 
mass}, i.e. massive enough to allow efficient gas cooling via roto-vibrational levels 
of H$_2$, $M>M_{\rm H_2}\approx 6\times 10^5\, [(1+z)/20]^{-3/2}\,\msun$ (virial 
temperatures $>2000\,$K); 
and b) Metal-poor halo stars (Population II) and globular clusters form 
in haloes above an {\it atomic cooling mass}, i.e. massive enough to allow efficient 
gas cooling and fragmentation via excitation of hydrogen Ly$\alpha$,
$M>M_{\rm H} \approx 10^8\, [(1+z)/10]^{-3/2}\,\msun$ (virial temperature $>10^4\,$K).

To test whether, at the present epoch, massive clusters contain a different 
fraction of high-$\sigma$ material than field galaxies, we have used 
a lower-resolution N-body simulation ($300^3$ particles in a 90 Mpc box, for 
a resolution of $10^{9}\,\msun$) in order to obtain a larger galaxy/cluster sample. We 
have selected ten parent hosts at $z=0$ with FOF masses close to $2\times 10^{11}\,\msun$, 
$10^{12}\,\msun$, $10^{13}\,\msun$, and $10^{14}\,\msun$, for a total of
fourty haloes. We have then marked FOF groups with more than 100 particles at redshifts
7.0, 4.3, 3.1 and 0.8, and determined the fractions of the virial mass of the parent
belonging to a fixed progenitor mass/redshift ($M, z$) pair. The values 
obtained are given in Table \ref{sigmafr}: the mass fractions are practically constant 
for all parents which host any of the selected progenitors. As we select parents
of lower masses, the number of parents hosting rarer (higher-$\sigma$) progenitors 
drops from 10 to 0 within about a decade in mass.


\section{Some applications}\label{App}

We have shown that the final distribution and kinematics of dark matter particles 
selected from early branches of the merger tree are systematically different 
than those of the parent halo as a whole. These properties are
also relevant for old stellar populations if these form predominantly
in early low-mass progenitor haloes, as stars behave essentially as collisionless 
systems just like the dark matter particles in our simulations. In the 
following we briefly discuss a number of possible applications.

\subsection{Remnants of the first stars}\label{PopIII}

Numerical simulations performed in the context of
hierarchical structure formation theories suggest that the first 
(Population III) stars may have formed out of
metal-free gas in dark matter minihaloes of mass above $6\times 10^5\,\msun$ 
(Abel et al. 2000; Bromm et al. 2002; Yoshida \etal 2003; Kuhlen \& Madau 
2005) condensing from rare high-$\sigma$ peaks of the primordial density fluctuation
field at $z>20$, and were likely very massive.
Barring any fine tuning of the initial mass function (IMF) of
Population III stars, intermediate-mass black holes (IMBHs) -- with masses above the
5--20$\,\msun$ range of known `stellar-mass' holes -- may be one of the 
inevitable endproduct of the first episodes of pregalactic star
formation (Madau \& Rees 2001). 

\begin{figure*}
\vskip 5.5 truein
\includegraphics{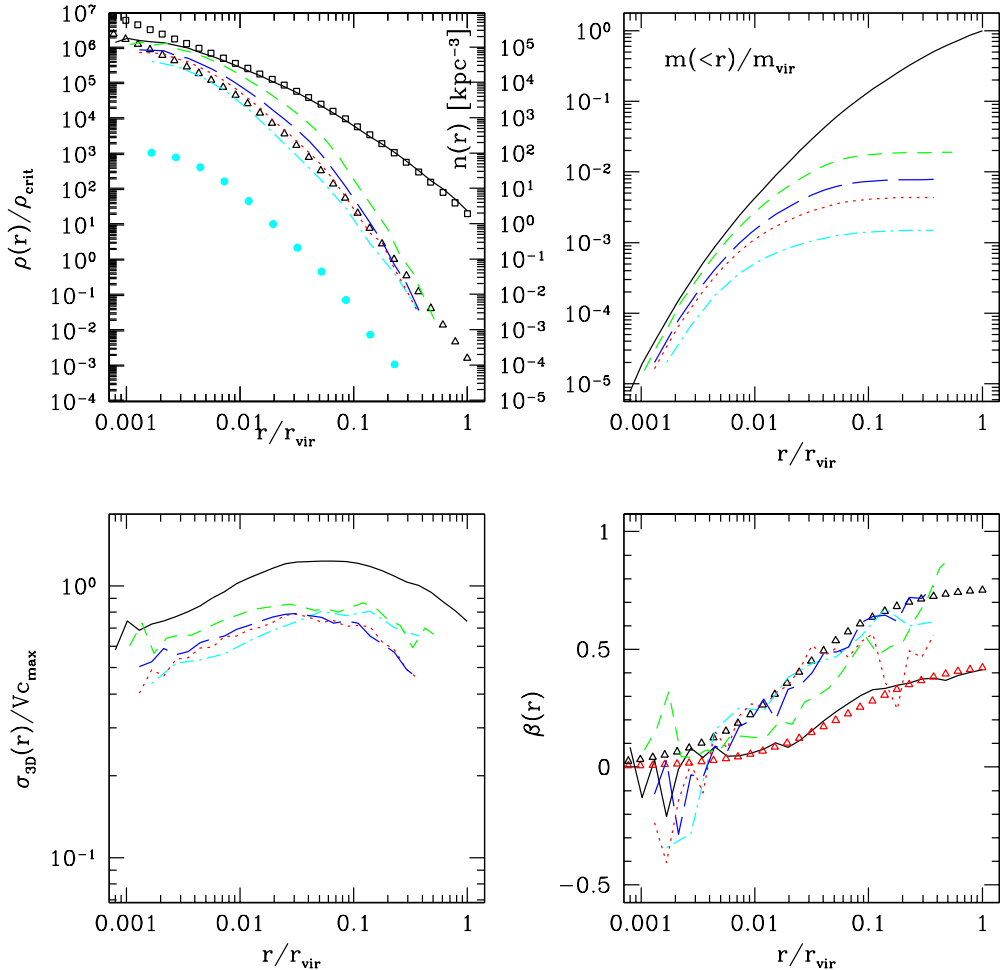}
\caption{Distribution of peaks above 3$\sigma$ in present-day galaxy-size haloes.
Line styles are as in Fig. \ref{avg1sig}. The circles in the upper left panel 
show the estimated mass and number density of wandering IMBHs assuming ${\cal N}=1$ 
seed hole for every $6\times 10^{5}$ solar masses of 3$\sigma$ progenitor halo material at 
$z_{0.9}=17.9$, and a final black hole mass of $1.5\times 10^4\,\msun$ (see text for 
details).} 
\label{avg3sig}
\end{figure*}

\begin{figure*}
\vskip 5.5 truein
\includegraphics{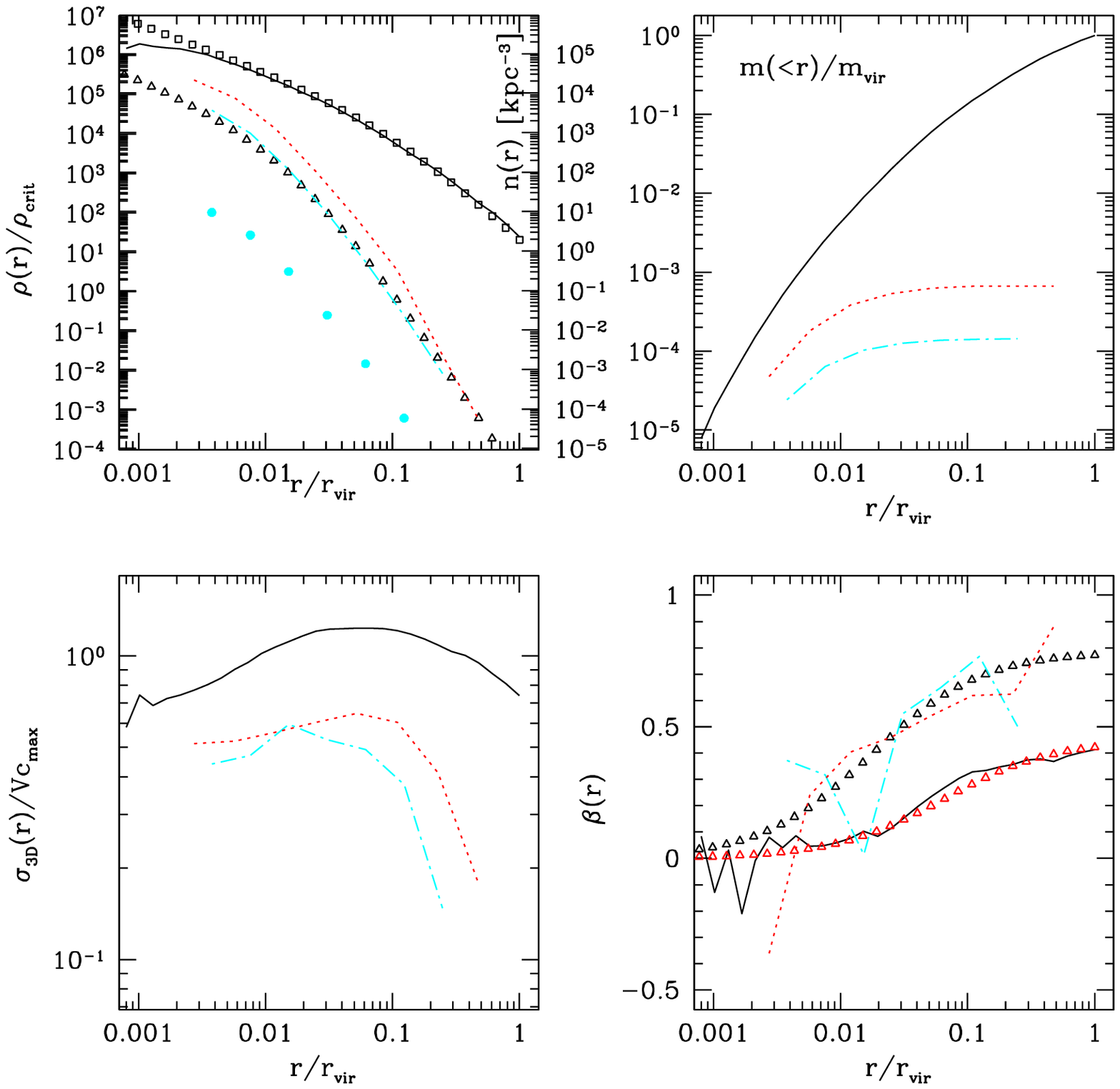}
\caption{Same as Fig. \ref{avg1sig} but for peaks above 3.5$\sigma$ and for 
Population III progenitors collapsing at $z_{0.9}=21.2$.
} 
\label{avg3.5sig}
\end{figure*}

Where do relic pregalactic IMBHs lurk in present-day galaxy halos?
To shed some light on this question, we have populated our 3$\sigma$ (3.5$\sigma$)
simulated progenitors at $z_{0.9}=17.9$ ($z_{0.9}=21.2$) with one seed IMBH for 
every $6\times 10^{5}$ solar masses of halo material. As discussed by Volonteri
et al. (2003), these IMBHs will undergo a variety of processes during the 
hierarchical buildup of larger and larger haloes, like gas accretion, binary hardening,
black hole mergers, triple interactions. While we neglect all of these effects here, 
our dark matter simulations do correctly model the bias in the 
formation sites, the accretion into larger haloes, and the competing effects of 
dynamical friction and tidal striping within larger potential wells, as these 
complicated dynamical processes are dominated by the dark haloes that 
host the black holes. Therefore, in our toy model, the distribution of 3 or 3.5$\sigma$ 
material at $z=0$ describes the properties of holes wandering through within today's 
galaxy haloes. The predicted IMBH number density and mass density profiles
are shown as circles in Figures \ref{avg3sig} and \ref{avg3.5sig}: 
the former may be regarded as an upper limit since we have neglected black hole mergers,
while the latter have been estimated assuming that these off-muclear black holes
have grown by accretion to a mean mass of $1.5\times 10^4\msun$, which is
a rough estimate obtained from Figure 14 of \citet{Volonteri2003}.

Depending on the IMF of Population III objects, some first-generation low-mass stars
may have survived until today. Their number density profile $n(r)$ within the Milky Way 
can again be read following the circles in Figure \ref{avg3sig}, under the assumption 
that ${\cal N}=1$ metal-free star survives for every $6\times 10^{5}$ solar masses of 
3$\sigma$ progenitor halo material at $z_{0.9}=17.9$. It is easy then to scale up the 
predicted value of $n(r)$ if ${\cal N}\gg 1$ such stars were to survive instead.   
On average, we find that about 1/3 of these remnants would lie today in the bulge, i.e.
in the inner 3 kpc. This fraction fluctuates between 24\% and 45\% in the four 
galaxies G0-G3. The density in the solar neighborhood is of order $\simeq 0.1\,{\cal 
N}\,$kpc$^{-3}$, three orders of magnitude lower than in the bulge. 
The number density of remnants would be lower and more concentrated toward the 
galactic center if Population III stars only formed within rarer 3.5$\sigma$ 
peaks (Fig. \ref{avg3.5sig}). On average, about 59\% of them would now lie 
within the bulge (range is 38\%-84\%) and the local number density would be only 
$\simeq 0.02\,{\cal N}\,$kpc$^{-3}$. Even for ${\cal N}=10-100$, this is an 
extremely small value, many orders of magnitude below the local number 
density of halo stars. {\it The above results suggest that  the very oldest stars and 
their remnants should be best searched for within the Milky Way bulge.}

\begin{figure*}
\vskip 5.5 truein
\includegraphics{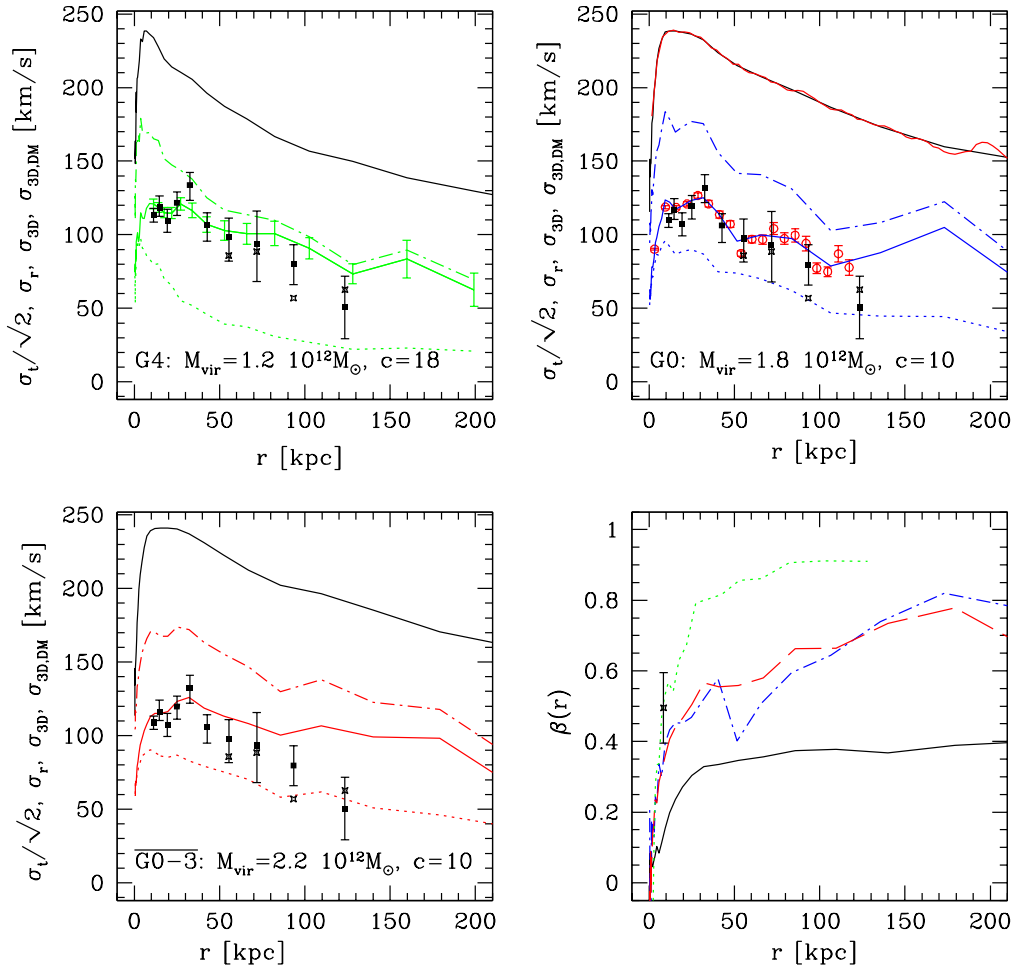}
\caption{Kinematics of our model stellar haloes compared with 
observations of Milky Way halo stars. The 
simulated CDM haloes are rescaled to give a local circular velocity of 220 km/s after 
taking baryonic contraction into account (see text for details). The lower right 
panel shows the predicted anisotropy profiles $\beta(r) = 1-\sigma^2_t(r)/2\sigma^2_r(r)$ 
for the G0-G3 average halo ({\it dashed line:} halo `stars', {\it solid line}: dark 
matter) and for halo stars in G0 ({\it dash-dotted}) and G4 ({\it dotted}). The 
data point is the local halo anisotropy from \protect\cite{Chiba2000}. The curves
in the other three panels show, from bottom to top, the `stellar'
$\sigma_t/\sqrt(2) \simeq\sigma_{\theta} \simeq \sigma_{\phi}$ ({\it dotted line}), 
$\sigma_r$ ({\it solid line}) and $\sigma_{\rm 3D}$ ({\it dash-dotted line}). The 
uppermost solid curve depicts the 3D velocity dispersion profile of the dark matter
(note that it is much larger and smoother than the stellar $\sigma_{\rm 3D}$). To 
illustrate the fluctuations in the stellar dispersions, 
$\sigma_{r}$ and $\sigma_{\rm 3D,DM}$ in halo G0 are plotted twice: once
using our usual choice of 30 logarithmic bins to the virial radius, and once using 
smaller linear bins (open circles with Poissonian error bars for $\sigma_{r}$,
another solid line for $\sigma_{\rm 3D,DM}$). The points are the measured $\sigma_r$
of Milky Way halo stars from \protect\cite{Battaglia2005}, with ({\it squares}) and
without ({\it stars}) the contribution from satellite galaxies. The
error bars reflect only the accuracy of the observations, while Poisson
noise (which causes similar, in some bins even larger, errors) is not included.
The declining dispersions near the center are characteristic
for halo containing {\it only} dark matter. In the Galaxy we
expect a flatter profile in the inner, baryon-dominated regions.
} 
\label{MWhalo}
\end{figure*}

\subsection{Stellar haloes}\label{stellarhalo}

Material from $>2.5\sigma$ peaks has today a density profile that is very similar 
to the stellar halo around the Milky Way \citep{Moore2005}. It contributes a few 
percent of the total virial mass, and therefore contains enough baryons to build up 
a $10^9\,\msun$ stellar halo with a reasonable star formation efficiency. 
The assumption of a common pregalactic origin between such a stellar component and
the surviving Local Group dwarf galaxies provides an additional constraint and 
allows us to determine the progenitor mass threshold/redshift pair which best
fit the data. From this argument \citet{Moore2005} identified hosts above $10^8\,\msun$
at $z=12$ as the progenitor haloes which the bulk of halo stars originally belonged to.

To check whether this simple model reproduces the kinematics as well as the radial 
distribution of halo stars, we have compared the predicted 
radial velocity dispersion profile with recent data from \citet{Battaglia2005} (see 
Fig. \ref{MWhalo}).  All haloes were rescaled (i.e. $r_{\rm vir}\to fr_{\rm vir}$, 
$V_c \to f V_c$, $M_{\rm vir}\to f^3 M_{\rm vir}$) to produce a local circular velocity of
220 km/s after taking into account the increase in circular velocity due
to baryonic contraction [$V_c(8.5\,{\rm kpc}) = 1.125\,V_{c,{\rm max,DM}}$] found 
by \citet{Maccio2005} for our halo G1. The rescaled virial masses are given in 
Figure \ref{MWhalo}. The predicted $\sigma_r(r)$ are close to the observed ones, 
especially for halo G0 and G4. Our model predicts a radial anisotropy that grows 
with radius (see lower right panel of Fig. \ref{MWhalo}): in the solar neighborhood 
it gives $\sigma_{\rm 3D}\simeq V_c$ and $\beta \simeq 0.4$, in good agreement 
with the observed values \citep{Chiba2000}. The declining velocity dispersion near the 
center is characteristic of haloes containing {\it only} dark matter. 
In the Galaxy we expect a flatter profile in the inner, baryon-dominated regions.

Both a high concentration ($c=18$), highly anisotropic halo like G4 and
a $c=10$ halo with smaller anisotropy like G0 fit the data equally well.
A detailed comparison requires knowledge of the tangential components of the 
velocity dispersion, in order to determine the concentration and virial mass of 
the Milky Way halo. The recently discovered stellar halo of M31 has a similar density
profile than that of our Galaxy \citep{Raja}. The line-of-sight velocity dispersion
observed in M31 is mostly due to tangential motions: preliminary results yield a constant 
$\sigma_t/\sqrt{2}$ of about 80 km/s for $50<r<150\,$ kpc \citep{Raja}, close 
to the value expected for the G0 scenario but much higher than that of the G4 model.
More data for M31 will hopefully soon become available, and many 
radial velocities of Milky Way halo stars will be obtained by surveys
like RAVE and SDSS-II. In the next decade the planned GAIA satellite will provide 
very detailed phase-space positions.

Another interesting feature of our model is that it predicts significant deviations 
from the simple radial velocity dispersion profiles expected for a smooth
stellar halo in an NFW potential (like the models in \citealt{Battaglia2005},
Fig. 4). These deviations are evident in the upper right panel of
Figure \ref{MWhalo} (but not in the lower left panel where we show an
average over four haloes).
\footnote{The sample from \citet{Battaglia2005} and our G4 outer stellar 
halo contain similar numbers of `stars'. Both
have relatively large Poisson noise and are consistent with smooth 
$\sigma_r(r)$ profiles. In haloes resolved with more particles like G0-G3 and D12, 
the `stellar' radial velocity dispersions are clearly inconsistent with smooth profiles.
}\, The fluctuations are not an artifact of
our analysis: with our usual choice of 30 logarithmic spherical bins
out to the virial radius [similar to the binning of the 
observed $\sigma_r(r)$] the fluctuations often extend over various bins with 
each bin containing thousands of `stars'. To show fluctuations
of $\sigma_r(r)$ on smaller scales we have binned `stars' in halo G0
using linear spherical bins, and have ploted the resulting  $\sigma_r(r)$
profile in the same radial range of the observations
(circles with error bars in the upper right panel of Fig. \ref{MWhalo}).
The linear bins contain at least 108 stars, the error bars represent Poisson noise.
We have also plotted $\sigma_{\rm 3D,DM}(r)$ of halo G0 twice using different bin sizes.
Note that the velocity dispersion of the dark matter halo as a whole (instead of 
the material belonging to the $>2.5\sigma$ $z=12$ subset) in our simulated galaxies
is generally a smooth function of radius, with small deviations becoming apparent only 
at $r\sim 200$ kpc. As fluctuations in the stellar $\sigma_r(r)$ are not caused by mass bound
to subhaloes (the mass fraction is much too small particularly within the inner halo), this
leaves tidal streams as the most probable cause of such deviations. The
entire dark halo is assembled from about $10^5$ resolved progenitors, the tidal 
debris of which are well mixed and smooth. In our model, the stellar halo 
is made up instead from only a few hundred building blocks (and a significant
fraction of it comes from the most massive ones), and is therefore 
expected to show some granularity. A more detailed study of tidal features in our model 
stellar haloes is left to a future paper.

\subsubsection{Comparison to recent related models for the formation of the stellar halo}

During the completion of this work two related models 
for the formation of the stellar halo (\citealt{Bullock2005}; \citealt{Abadi2005})
came to our attention. Both studies are more detailed than our in the sense that they try
to model the effect of star formation and chemical evolution. 
\cite{Bullock2005} use a semi-analytic approach combined with N-body 
simulations of satellite 
disruption in an external, growing galaxy potential. Like in this work the number of 
luminous satellites and the mass of the final stellar halo is adjusted to match 
the observations 
by assuming that reionisation suppresses star formation in later-forming small progenitors.
Our stellar haloes, however, seem to be more concentrated than those of \cite{Bullock2005}.
This may due to the fact that we only trace old stellar populations, 
while \cite{Bullock2005} also allow for more recent episodes of
star formation in their protogalactic building blocks. They also model the
disruption of satellites assuming an external spherical potential: this is 
probably a good approximation to recent accretion events which build up
the outer stellar halo. The inner stellar halo, however, is built up through a series
of early, major mergers: these cause rapid potential fluctuations (violent relaxation) 
which are not accounted for in the spherical potential approximation. This limitation 
was also pointed out by \cite{Bullock2005}.

The cosmological SPH simulations of \cite{Abadi2005}, while not including any global 
mechanism for suppressing star formation in later-forming small progenitors, do
lead to similar stellar halo profiles and kinematics as our model. They seem
to produce, however, far too many halo stars: recent SPH simulations at higher 
resolution and without strong feedback effects also overpredict the number of satellites 
by a large factor \cite{Maccio2005} and tend to produce even more halo stars. 
The similarity of the profile shape and kinematics of stellar haloes in all 
these models suggests that they are a robust and generic 
outcome of hierarchical structure formation (see also \citealt{Hansen2004}; 
\citealt{Dekel2005}) and do not depend on the detailed formation history. Note that 
this does not contradict our argument about the possibility to use the radial 
extent of collisionless tracer populations to learn about their formation epoch 
and sites. It just illustrates how this arguments is limited to fossil records from 
very high redshift ($z>8$, see also \S~\ref{region}). At
lower redshifts the typical haloes are larger and the exact star formation sites must 
be taken into account. Efficient cooling (and/or numerical losses of gas angular momentum 
to the dark matter in under-resolved progenitors) can produce a population of stars at some
intermediate redshift with a distribution that is similar to much older populations
from early high-$\sigma$ peaks. By $z=0$ the two subsets would have very similar
concentrations and kinematics, which might explain the similar results found here 
and in \cite{Abadi2005}.

\subsection{Metal-poor globular clusters}

The clustering properties of metal-poor globular clusters contain 
clues on their formation sites and 
the epoch when star formation was suppressed by feedback
processes (e.g. reionisation, supernova-driven winds) in low-mass haloes.
In the Milky Way metal-poor globular clusters follow the same
radial profile as halo stars, suggesting within the framework of our model a 
common origin within early 2.5$\sigma$ peak progenitors at $z\approx 12$.
Observations of the distribution of globular clusters within present-day haloes 
of different masses could provide information on feedback effects as a 
function of environment \citep{Moore2005}. The suppression of globular cluster 
formation at some early epoch may also explain the bimodality observed in cluster
metallicities (e.g. \citealt{Strader2005}).
The widely used assumption that globular cluster formation
is a fair tracer of star formation, combined with the suppression of 
the formation of metal-poor globulars after reionisation, imply 
that the amount of high-$\sigma$ material in a halo is proportional
to the number of metal-poor globular clusters. From the results in \S~\ref{fraction} 
it follows then that a simple universal reionisation epoch would lead to 
a constant abundance of metal-poor globular clusters per virial mass.
Deviations from this simplest case may provide information about the local
reionisation epoch, and whether regions with more (less) metal-poor globulars
per virial mass were reionised later (earlier) (see \citealt{Moore2005}).

\subsection{Elliptical galaxies}

The projected luminosity profiles of elliptical galaxies follow the R$^{1/4}$ law
and resemble rescaled versions of projected CDM halo density profiles 
(\citealt{Lokas2001}; \citealt{Merritt2005}),
similar to the $z=0$ profiles of our high-$\sigma$ subsets.
Ordinary elliptical galaxies may form by multiple mergers or by the major merger 
of two disk galaxies (e.g. \citealt{Dekel2005}).
Early formation ($z \simeq 6$) in a large number of progenitors 
followed by a series of gas-poor, 
essentially collisionless mergers 
could build up the giant ellipticals (cD) in the centers
of galaxy clusters. These mergers might undo some of the early dissipational
contraction of the total mass distribution \citep{Gao2004}.

\begin{figure*}
\vskip 5.5 truein
\includegraphics{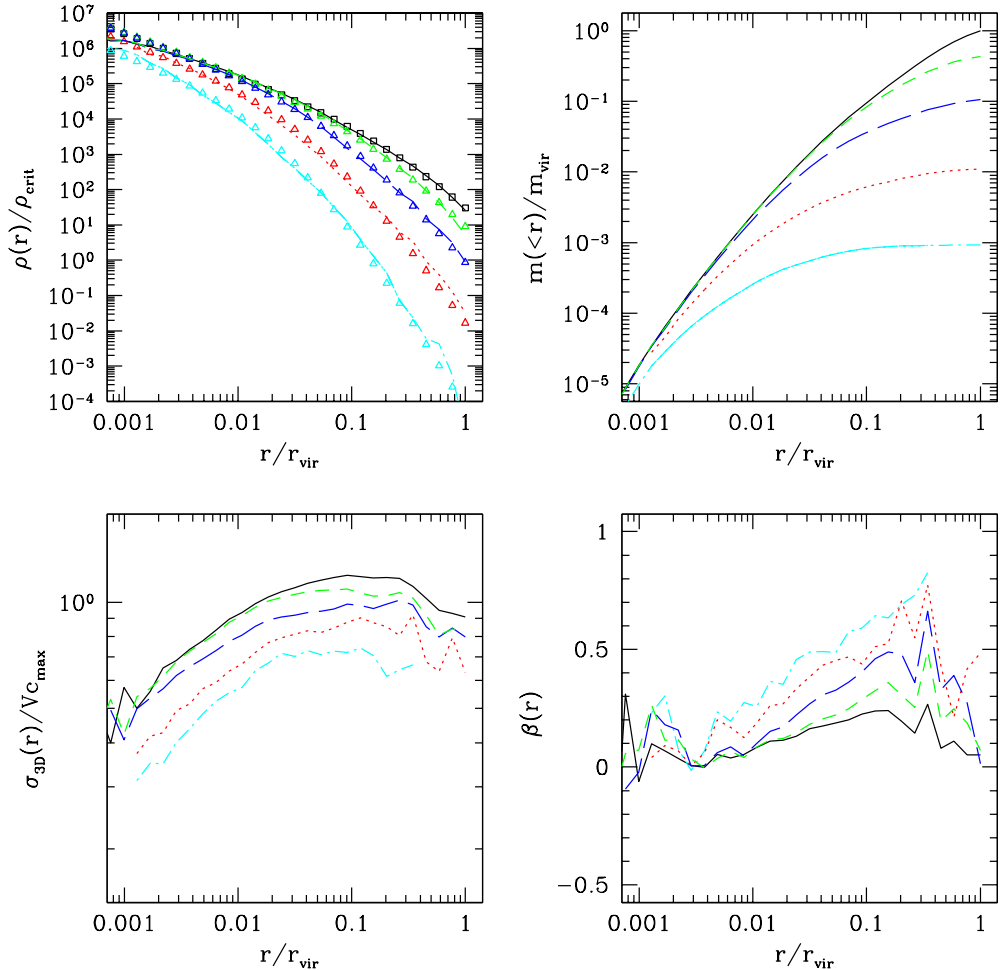}
\caption{Distribution of peaks above 1 (short-dashed),
2 ({\it long-dashed}), 3 ({\it dotted}) and 4$\sigma$ ({\it dash-dotted})
in a present-day cluster-size halo ($M_{\rm vir}=3.1 \times10^{14} \msun$).
The redshift and minimum mass pairs used are ($z=1.6$, $7.4\times 10^{9}\msun$) for
the 1$\sigma$ peaks, (4.4, $6.3\times 10^{9}\msun$) for the 2$\sigma$,
(7.4, $4.3\times 10^{9}\msun$) for the 3$\sigma$, and
(8.7, $1.8\times 10^{10}\msun$) for the 4$\sigma$ peaks.
The triangles in the upper left panel show the fitting function (\ref{emp}).
}
\label{allsig}
\end{figure*}

\begin{figure*}
\vskip 5.5 truein
\includegraphics{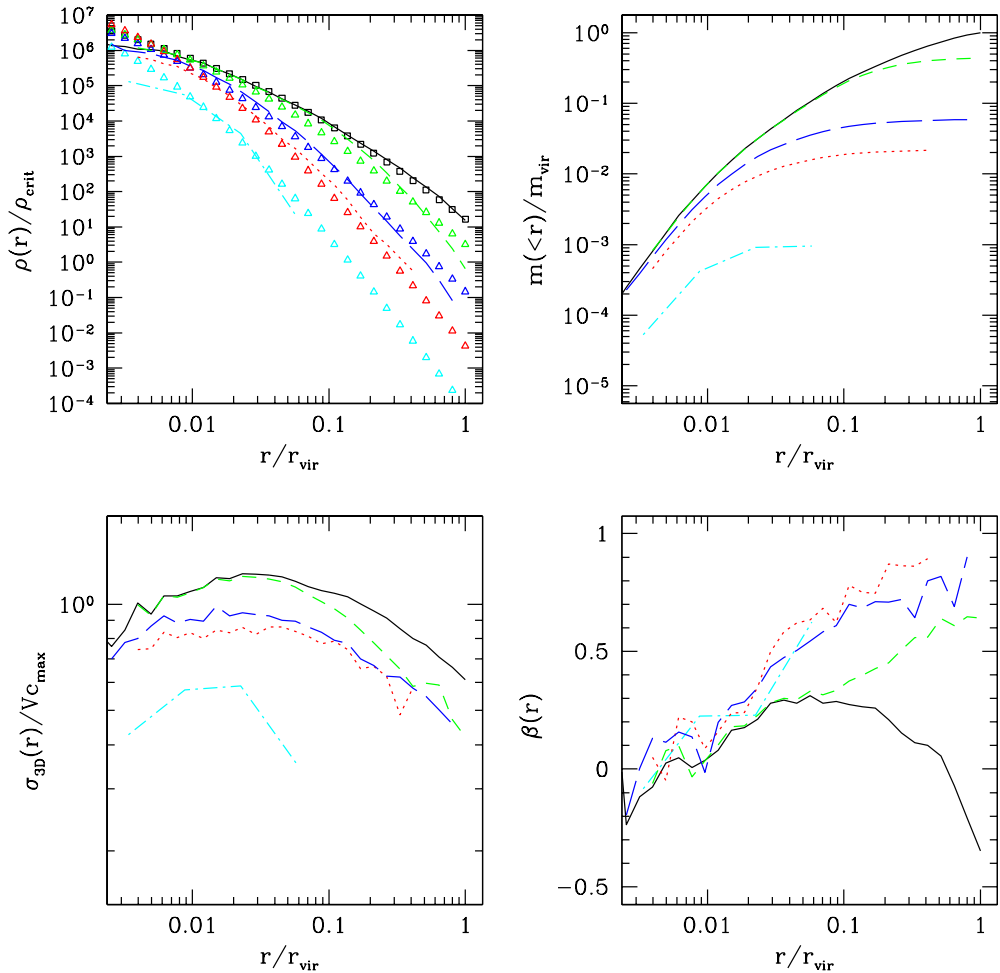}
\caption{Same as Fig. {\ref{avg1sig}} for peaks above 1 ({\it short-dashed}),
2 ({\it long-dashed}), 3 ({\it dotted}) and 3.5$\sigma$ ({\it dash-dotted})
in a small galaxy halo ($M_{\rm vir}=10^{11} \msun$) at $z=0$.
The redshift and minimum mass pairs used are ($z=2.4$, $4.8\times 10^{8}\msun$) for
the 1$\sigma$ peaks, (8.7, $7.6\times 10^{6}\msun$) for the 2$\sigma$,
(10.5, $2.2\times 10^{8}\msun$) for the 3$\sigma$, and
(13.7, $ 4.9\times 10^{7}\msun$) for the 3.5$\sigma$ peaks.
The triangles in the upper left panel show the fitting function (\ref{emp}).
}
\label{dallsig}
\end{figure*}

Figure \ref{allsig} shows the distribution of material belonging to different 
$\sigma$ peaks in a $z=0$ cluster halo.
The inner cluster regions are entirely made up of material from $>2\sigma$ peaks,
and this selection contributes about 10\% to the cluster virial mass. 
The central brightest galaxies in SDSS clusters have an average effective radius
of $R_e \simeq 20\,$kpc and presumably about 10\% of the clusters luminosity \citep{Zibetti2005}. 
Our $2\sigma$ subsets contain enough mass to account for the observed cD luminosities
but they are more extended. If the cD stars formed early
in the inner parts of all progenitor haloes above 2$\sigma$ then they would be
found closer to the cluster center today (see Section \ref{region}). 
A simple way to select the inner parts
of 2$\sigma$ progenitors is to mark higher-$\sigma$ peaks at earlier epochs.
Our 3$\sigma$ selections have a realistic
effective radius of about $20 $ kpc. Therefore we predict the 
kinematics of the average brightest cluster galaxies to follow
the 3$\sigma$ selections shown in Figure (\ref{allsig}) if (or where) the
real cluster potentials are similar to the uncontracted potentials
formed in dissipationless cosmological simulations (as 
suggested by \citealt{Gao2004}).

The anisotropy of a $z=0$ subset depends 
mostly on its density profile shape (or $\nu$)
and not on the selection redshift or
the number of mergers it has undergone (Figures \ref{avg1sig} to \ref{avg3.5sig})
A few mergers are enough to
establish the general $\beta(r)$ profile (\ref{bemp}) and the slope --
anisotropy relation (Section \ref{slope-beta}). 
This agrees with
\citet{Moore2004} who find $\beta \simeq 0.5$ after only one low angular
momentum merger of initially isotropic CDM haloes (see also \citealt{Hansen2004}).
\cite{Dekel2005} also find a similar $\beta(r)$ profiles after only one merger,
which even starts with disk galaxies, i.e. nearly circular initial
orbits for the stars 
(see their $\beta$ for the old stars, Fig. 1 of \citealt{Dekel2005}).
Their average anisotropy at the half mass radius of the old stellar
component is $\beta=0.3$, which is lower than our average of $\beta=0.45$,
but considering the very different initial conditions of the stars 
and the larger numbers of mergers in our cosmological context the results
are surprisingly similar. Therefore we argue that our collisionless
results are relevant both  for giant elliptical galaxies which
formed in a series of dissipation-less mergers \citep{Gao2004} and also for
the old stars in smaller ellipticals.

The general relation between density pofile slope and anisotropy 
for high sigma subsets $\beta(d \log \rho / d \log r) \simeq
 -0.23(1.2 + d \log \rho / d \log r )$
can be applied to elliptical galaxies to infer $\beta(r)$ from the
observed stellar profile. This is useful to obtain dynamical mass 
estimates for elliptical galaxies, where one 
has to assume some $\beta(r)$ \citep{Mamon2005} and
it is probably the best guess before realistic, cosmological, hydrodynamical
simulations of elliptical galaxy formation become available. 

\section{Conclusions}\label{last}

We have used high-resolution cosmological N-body simulations to trace the 
spatial distribution and kinematics in present-day CDM haloes of collisionless 
material that originally belonged to selected branches of the merger hierarchy. 
Our main results can be summarized as follows:

\begin{enumerate}
\item Hierarchical merging does not efficiently mix haloes up.
The final distribution of dark matter particles retains a memory of when and
where they collapsed initially, allowing a unique test of the popular
bottom-up paradigm for the formation of cosmic structure.

\item Today's distribution and kinematics of halo substructure that formed 
at very high redshift depends mostly on the rarity of the primordial density 
fluctuations they correspond to. For example, the $z=0$ 
density, shape, anisotropy, and velocity dispersion profiles of 
material originating from early 3$\sigma$ peaks is independent of the 
redshift/minimum mass pair used to select it. The mass fractions within the parent
host, however, grow if such particles are choosen from lower redshift progenitors 
at a higher mass threshold.

\item High-$\sigma$ material should be looked for close to the center of massive 
parent haloes. 
The concentration and outer slope of the mass density profile are larger for the 
rarer peaks. 

\item The anisotropy increases faster with radius for material originating 
within rarer peaks (eq. \ref{bemp}). The parameter $\beta(r)$ steepens with increasing
$\nu$ at the same rate as the half-mass radius shrinks, i.e. we generally find 
$\beta(r_{1/2}) \simeq 0.45$.

\item Velocity dispersions are lower for material from the rarer peaks, and 
particle orbits are more radial. While the average velocity dispersion profiles
agree with stationary solutions to the Jeans equation (\ref{je}), the profiles
of individual haloes show significant structure and are inconsistent with smooth 
$\sigma(r)$ solutions (see Fig. \ref{MWhalo}).

\item The high-$\sigma$ subsets have much more elongated shapes than their host haloes.
For 3$\sigma$  material the mean $c/a$ is about 0.35, while present-day galaxy-size 
haloes have a mean of 0.6.

\item These properties do not depend on which regions within early ($z>8$) progenitors 
are marked and traced to the present time. Particles originating within the cores
of high-$\sigma$ progenitors are distributed at $z=0$ in the same way as the entire
virial mass of these early structures. This is not true at later epochs, where 
the marking of only the densest progenitor regions results in more concentrated
density profiles (e.g. the selection of only the densest 10\% of the mass of 
progenitor halos above 1$\sigma$ at $z=1.6$ yields a present-day distribution 
similar to that of all particles above 2$\sigma$).

\item If the first stars form at early epochs in peaks above 3.5$\sigma$, then 
half of their remnants should be found in the bulge, within 3 kpc of the galactic 
center. In the solar neighborhood the density of such very old population is 1000 
times lower than in the bulge. Also, their characteristic velocities are 2.5 times 
lower than those of dark matter particles, and their orbits more radial ($\beta 
\simeq 0.5$).

\item The radial profile of the stellar halo and metal-poor globular clusters of 
the Milky Way suggest that these components formed in rare early peaks above 
2.5$\sigma$ at redshift above 10. Typical outer halo objects have radial orbits, 
and become isotropic near the galactic center.

\item Radial orbits are a general outcome for any concentrated stellar component 
assembled through gas-poor mergers. The anisotropy parameter $\beta(r,\nu)$ correlates 
well with $\rho(r,\nu)$ and is not sensitive to the detailed assembly history of 
such a component. 

\end{enumerate}

The applications discussed above should be taken as a first attempt at 
predicting the spatial distribution and kinematics in present-day galaxies 
of objects that formed within early protogalactic systems, all in the
context of hierarchical structure formation theories. Some interesting directions 
for future work may include, e.g., combining spatial information on Population III
remnants with semi-analytic prescriptions for the growth and dynamical evolution of 
black holes, as well as using chemical evolution models to translate the age gradient
for stellar haloes found here into metal abundance gradients. The techniques
presented here also allow to study the properties of stellar streams in realistic, 
triaxial, clumpy CDM galaxy haloes.

\section*{Acknowledgments}
It is a pleasure to thank G. Battaglia for
providing us with stellar halo data in electronic form.
We are grateful to G. Battaglia, M. Beasley, A. Faltenbacher, M. Kuhlen and M. 
Volonteri for helpful suggestions and discussions.
All computations were performed on the zBox
supercomputer at the University of Zurich. Support for this work was provided by 
NASA grants NAG5-11513 and NNG04GK85G (P.M.), NSF grant AST-0205738 (P.M.), and by the 
Swiss National Science Foundation (J.D.). P.M. also acknowledges support from 
the Alexander von Humboldt Foundation. Part of this research was carried out at 
the Kavli Institute for Theoretical Physics, UC Santa Barbara, under NSF Grant 
No. PHY99-07949.

\bsp
\label{lastpage}
\end{document}